\documentclass[11pt]{article}
\usepackage[colorlinks, linkcolor=blue, citecolor=blue, urlcolor=blue, bookmarks, bookmarksnumbered]{hyperref}
\usepackage{amsmath, amssymb, mathrsfs}
\usepackage{cite, xfrac}

\newcommand\be{\begin{equation}}
\newcommand\ee{\end{equation}}
\newcommand{\bm}[1]{\boldsymbol{#1}}
\newcommand{\pa}{\partial}

\newcommand{\rd}{\mathrm{d}}
\newcommand{\rD}{\mathrm{D}}
\newcommand{\K}{\mathbb{K}}
\newcommand{\CN}{\mathcal{N}}
\newcommand{\Q}{\mathbb{P}} % Why is \Q -> \mathbb{P} ???

\newcommand{\X}{\mathbb{X}}
\newcommand{\sfrc}[2]{\sfrac{#1\!}{#2}}
\newcommand{\define}{\mathrel{\overset{\scriptscriptstyle\rm def}=}}
\DeclareMathOperator{\Tr}{Tr}
\DeclareMathOperator{\Vol}{Vol}

\def\MeV{\text{MeV}}

\def\2#1{{\color{blue}#1}}

\def\define{\buildrel{\rm def}\over=}

 \allowdisplaybreaks
 \numberwithin{equation}{section}
 
   \parskip\medskipamount     \lineskip=0pt
\thicklines \footnotesep=2mm
\begin{document}

\thispagestyle{empty}
\vglue5mm
\begin{center}
{\LARGE\bf
  Nambu's Dynamics, Strings, and\\*[1mm] 
  Gravitized Quantum Theory}\\*[5mm]
{\bf Tristan H\"ubsch}$^*$ and {\bf Djordje Minic}$^\dag$\\*[1mm]
{$^*$\slshape Department of Physics and Astronomy, Howard University\\* Washington, DC, 20059, USA; \texttt{thubsch@howard.edu}}\\*
{$^\dag$\slshape Department of Physics, Virginia Tech,\\*
Blacksburg, VA 24061, USA; \texttt{dminic@vt.edu}}\\*[5mm]

{\bf ABSTRACT}\\*[2mm]
\parbox{142mm}{\addtolength{\baselineskip}{-3pt}\parindent=2pc\noindent
In this article we present a new view on string theory that is rooted in quantum foundations extended to the domain of quantum gravity in the form of gravitized quantum theory, in which the canonical probability measure captured by the Born rule of quantum theory, turns out to be dynamical and contextual. We claim that this gravitization of quantum theory is the essence of string theory. String theory, according to this view, originates in foundations of quantum theory and Nambu's dynamics, which features in the non-perturbative definition of string theory in terms of a T-duality covariant and non-commutative formulation, called the metastring. Nambu's dynamics and the associated cubic and hypercubic objects also appear as the natural language for exploring the new observables and phenomena implied by a dynamical Born rule, including triple and higher order quantum interference, that can be investigated in experiments involving massive quantum probes, and that are also needed for direct empirical tests of the Born rule, a cornerstone of quantum theory.
}
\vspace{5mm}

\begin{minipage}{142mm}\small
  \baselineskip=10pt plus1pt minus 1pt
  \setcounter{tocdepth}{3} % reduce to 2 ?
  \tableofcontents
\end{minipage}
\end{center}

\vspace{2mm}
%\clearpage

%============================================================================
\section{Introduction and Overview: What is  String Theory?}
\label{s:IRS}

In 1925 Werner Heisenberg published~\cite{Heisenberg:1925zz} what is considered the first paper on quantum mechanics (later formalized and better understood by
Born and Jordan~\cite{Born:1925mph}, Born, Heisenberg and Jordan~\cite{Born:1926uzf} and Dirac~\cite{Dirac:1925jy}) in which new matrix observables were introduced into physics (capturing the fundamental aspects of atomic spectroscopy and translating the observed frequencies of spectral lines into the rule for matrix multiplication)
while keeping the equations of motion known from classical Hamiltonian dynamics the same.
As realized by Dirac, the classical Poisson equations get replaced by Born--Heisenberg--Jordan equations of motion in which the Poisson bracket is replaced by the commutator of matrices, and, as realized by Born, the fundamental commutator between matrix configuration variables ($q^a$) and their conjugate momenta ($p_a$), $[q^a, p_b]= i \hbar \delta^a_b $ is postulated (thus introducing the Planck constant into quantum mechanics in its matrix mechanics form). This non-commutativity implies that the theory is intrinsically probabilistic and that the relevant probability measure is non-classical. Thus quantum theory is an intrinsically probabilistic theory (with interfering probabilities) in which probability measure (the Born rule~\cite{Born:1926yhp}) is consistent with equations of motion. (We will later discuss the uniqueness of the Born rule and the possibility of a more general quantum theory with dynamical and contextual probabilities.)

In 1973 Yoichiro Nambu~\cite{Nambu:1973qe} introduced his ingenious generalized Hamiltonian dynamics, with an immediate question of its quantization (for a nice review consult the recent book by Tamiaki Yoneya~\cite{Yoneya:2026Int}).
Nambu's dynamics~\cite{Nambu:1973qe} is defined in terms of a natural generalization of the Poisson bracket, and is represented by a triple and higher order Nambu bracket that generates volume and hypervolume diffeomorphisms (thus generalizing area preserving diffeomorphisms of the Poisson bracket). The Nambu bracket acts linearly, but it is characterized by multiple (two or more) Hamiltonians. There are many superintegrable classical systems~\cite{Takhtajan:1993vr} (such as the asymmetric top, Nahm's systems, strings/membranes) which can be written in terms of the Nambu bracket, but the empirical analog of atomic spectroscopy from the historical development of quantum mechanics has been lacking (even though we will suggest that such an analog exists in the later sections of this paper).

Given the structure of canonical quantum mechanics in its Born--Heisenberg--Jordan form, the question of the quantization of Nambu's generalized dynamics became an important mathematical physics question, but the physical application of Nambu's dynamics remained obscure. (Nambu himself had an intuition that his generalized dynamics would be important for string theory, and in particular, the QCD string~\cite{Nambu:1973qe}.) As far as generalizations of quantum theory go, most physicists are very skeptical on this topic, as there is no indication that quantum theory fails in any of the existing experiments, and it is not clear that any proposed generalizations are consistent~\cite{Aaronson:2004ukm}. Of course, the interpretational questions in quantum theory are still being discussed, studies of quantum information, complexity and entanglement (``the second quantum revolution"~\cite{Dowling:2002jjw, Acin:2017gyd}) are thriving, and there is still the problem of precisely understanding the quantum-to-classical transition; this includes the problem of measurement, as well as understanding the quantum properties of ``macroscopic'' objects, what qualifies an object as ``macroscopic,'' and where and how a demarcation from the underlying ``microscopic'' dynamics emerges~\cite{Zurek:2003zz}. There is the frontier of quantum properties of gravity, which are only beginning to be explored experimentally~\cite{Bose:2017nin, Marletto:2017kzi, Penrose:2014nha, Zurek:2020ukz, Berglund:2023vrm}. At the moment quantum theory looks very robust, and one always wonders whether any of the proposed generalizations will prove to be lasting: Past proposals have problems with causality, or conservation laws, or cloning, or negative probabilities, or consistent formulation of composite systems, etc (see for example~\cite{Weinberg:1989Tes, Polchinski:1991Wei, Mielnik:2000ej, Aaronson:2004ukm}).
So, can a quantum formulation of the Nambu dynamics fare any better? 

In this paper we want to claim that this is the case and that
this generalized view of quantum theory, called ``gravitized quantum theory,'' in which the probability measure (the famous Born rule of canonical quantum theory) is dynamical and 
contextual~\cite{Minic:2003en, Minic:2002pd, Minic:2003nx, Minic:2004rj, Jejjala:2007rn, Freidel:2014qna, Minic:2020oho, Hubsch:2024agh}, is what lies in the foundations of string theory. String theory we argue originates in foundations of quantum theory, although it historically originated in the problem of strong interactions, only to be transformed into a unified theory of perturbative quantum gravity and (beyond) the Standard Model matter~\cite{Rickles:2014fha, Hubsch:2026The}. We also argue that Nambu's dynamics (which features in the non-perturbative definition of string theory in terms of a T-duality covariant and non-commutative formulation, called the metastring) as well as the associated cubic and hypercubic algebraic objects appear as the natural language for exploring the phenomena implied by a dynamical Born rule --- such as triple and higher order quantum interference, which can be studied in experiments involving massive quantum probes~\cite{Berglund:2023vrm, Berglund:2025qyo, Berglund:2025yoy}. Moreover, such experimental probes of triple interference are needed as direct empirical tests of the Born rule~\cite{Sorkin:1994dt, Bhatta:2026wuz}, one of the cornerstones of quantum theory, on which quantum technology~\cite{Dowling:2002jjw, Acin:2017gyd}, currently under development, is based.

Coming back to the the formal development of Nambu's generalized dynamics, in the 1980s, various physicists working on the dynamics of supermembranes (see, for example~\cite{Bergshoeff:2025haj} for a recent discussion of the history of membrane theory) and higher dimensional so-called p-branes
(especially in relation to supergravity theories in various dimensions~\cite{Bergshoeff:1987cm}) realized the connection between the world-volume actions of p-branes and Nambu's generalized dynamics~\cite{Bergshoeff:1988hw}. This prescient work also suggested that the quantization of Nambu's bracket would be useful for quantization of membranes and p-branes, in general. In the 1990s, after the advent of the idea of M-theory in string theory~\cite{Polchinski:1998rq, Polchinski:1998rr, rBZ-StrTh}, the mysterious formal unification of all superstring theories and their dualities~\cite{Polchinski:1998rr}, as well as various D-brane and p-brane like objects~\cite{Polchinski:1998rq, Polchinski:1998rr, rBZ-StrTh}, including the famous 11 dimensional supergravity~\cite{Cremmer:1978km}, with a direct connection to the quantum theory of supermembranes~\cite{deWit:1988wri}, it became apparent that Nambu's bracket and its quantization should be important (at least, formally) for a more profound understanding of the quantum structure of M-theory. In particular, a matrix model version of the light-cone supermembrane in 11 dimensions~\cite{deWit:1988wri}, known as M(atrix) theory, or the BFSS matrix model~\cite{Banks:1996vh}, or the theory of $N$ interacting $D0$-branes, was suggested as a light-cone formulation of M-theory, with important ramifications for the deep structure of string theory and its non-perturbative formulation, especially the IIA string. (Another important matrix model was suggested as a non-perturbative formulation of the IIB string~\cite{Ishibashi:1996xs}.) The Nambu triple bracket, as well as cubic (rank-3) matrices, were used in~\cite{Awata:1999dz}, in attempting for formulate a covariant and fully quantum version of M(atrix) theory~\cite{Banks:1996vh}, as a non-perturbative formulation of string/M theory. (However, no connection of such cubic matrices to triple interference was discussed, and this was only recently pointed out in~\cite{Bhatta:2026wuz}.) Later research led into very important, but different, formulations of membrane theories (Bagger--Lambert--Gustavsson~\cite{Bagger:2006sk, Bagger:2007jr, Gustavsson:2007vu} and ABJM~\cite{Aharony:2008ug} formulations) that did invoke higher order algebraic structures related to the Nambu bracket, but definitely {\it not} the idea of a possible gravitization of quantum theory needed for elucidating the physical principle behind the foundations of a non-perturbative formulation of string and M-theory and their explicit empirical signatures.

The formal question of quantization of Nambu's dynamics though remained opened (with some abstract solutions being proposed, such as Zariski quantization~\cite{Dito:1996xr, Sato:2012ti}, which was then used in a proposal for a covariant formulation of M-theory~\cite{Minic:1999js}). Some researchers~\cite{Curtright:2002hm, Zachos:2003md, Curtright:2003fn, Curtright:2003wj, Zachos:2004zc}, suggested that one could quantize Nambu dynamics by quantizing superintegrable dynamical systems, following the usual quantization rules (such as phase space quantization~\cite{Curtright:2011vw}). Still, with all these important developments and insights, the physical meaning of Nambu's dynamics and its quantization remained elusive. However, most recently Nambu's dynamics and its quantum formulation have been related to some foundations questions in quantum theory, and in particular, the empirical searches for the validity of the Born rule~\cite{Sorkin:1994dt}, and its proposed dynamical and contextual generalization in the realm of quantum gravity~\cite{Minic:2002pd, Minic:2003en, Minic:2003nx, Minic:2004rj, Freidel:2014qna, Berglund:2022skk, Berglund:2023vrm, Berglund:2025qyo, Berglund:2025yoy}. In particular, in the recent work~\cite{Bhatta:2026wuz} it has been shown that Nambu's dynamics of ``density cubes'' (rank-three objects that generalize the familiar density matrices of quantum theory) leads to the phenomenon of intrinsic triple interference~\cite{Berglund:2023vrm}, which is used to quantify empirical probes of the validity of (or %any % if the first is rank-four or higher
departure from) the Born rule~\cite{Sorkin:1994dt}. This also ties to the recent understanding of quantum aspects of measurement in the context of gravity, which directly relate to a dynamical and contextual nature of the Born rule in quantum gravity~\cite{Berglund:2025qyo, Berglund:2025yoy}.
(Somewhat earlier, it was realized that a version of Nambu quantum theory~\cite{Minic:2002pd}, closely related to Nambu's dynamics, can be used in the context of neutrino physics~\cite{Minic:2020zjb} in relation to non-electromagnetic experimental tests of the Born rule~\cite{Huber:2021xpx, Bhatta:2023biy}.)

What we wish to point out in this paper is the appearance of the Nambu bracket in a proposed non-perturbative formulation of metastring theory, a chiral phase-space-like and T-duality covariant formulation of string theory. This also points to a fundamental realization that string theory originates in quantum foundations, which is more apparent in the metastring formulation~\cite{Freidel:2013zga, Freidel:2014qna, Freidel:2015pka, Freidel:2015uug, Freidel:2016pls, Freidel:2017nhg, Freidel:2017wst, Freidel:2017xsi} than in the (by now textbook, standard) ``canonical'' one~\cite{Polchinski:1998rq, Polchinski:1998rr, rBZ-StrTh}. This new insight (perhaps not surprising, given the infinite number of conserved quantities of the free string which can be used as Hamiltonians of Nambu's generalized dynamics) also illuminates the origins of M-theory and its relation to the new fundamental observables in quantum gravity, as exemplified by the triple and higher order interference phenomena involving experiments with massive quantum probes~\cite{Berglund:2023vrm}. We also comment about the possible relevance of this new understanding of string theory to other outstanding problems in fundamental physics, such as the origin of visible and invisible (dark) matter as well as dark energy~\cite{Berglund:2020qcu, Freidel:2021wpl, Freidel:2022ryr, Freidel:2023ytq,  Minic:2023oty, Berglund:2023gur, Hubsch:2024agh, Hubsch:2026lot}.

The paper is organized as follows: In Section~\ref{s:Nambu2Strings} we first give an overview of Nambu's dynamics and it's relation to string and M(atrix) theory (M-theory in the light-cone gauge), via the connection between the triple Nambu bracket and the eleven dimensional supermembrane. In Section~\ref{s:Strings2Nambu} we then review the metastring formulation of string theory, which is T-duality covariant, phase-space-like and intrinsically non-commutative. In Section~\ref{s:Nambu2GQTh} we finally propose a non-perturbative formulation of the metastring with an explicit connection to the triple Nambu bracket: In view of our recent results~\cite{Bhatta:2026wuz}, this provides the main claim of the paper, to wit, that string theory is a gravitized quantum theory and as such endowed with observables more general than S-matrices (such as S-cubes and S-hypercubes), providing a direct experimental probe represented by the phenomenon of intrinsic triple interference of massive quantum objects. Section~\ref{s:Coda} concludes with a brief summary of our salient points.
So as to keep the presentation self-contained in summarizing various topics presented in this paper, we closely follow and build upon our previous papers~\cite{Jejjala:2007rn, Minic:1999js, Freidel:2013zga, Minic:2020oho, Bhatta:2026wuz}.

%============================================================================
\section{From Nambu's Generalized Dynamics to Strings}
\label{s:Nambu2Strings}
\subsection{Nambu's Dynamics}
In Nambu's dynamics~\cite{Nambu:1973qe} the classical equation of motion based on the Poisson bracket 
\be
\frac{\rd }{\rd t} F = \{ F, H \},
\ee
where the Poisson bracket on phase space is defined as
\begin{equation}
\{A,B\} = \sum_{ij}\varepsilon_{ij}\,
\dfrac{\partial A}{\partial q_i}
\dfrac{\partial B}{\partial q_j}
\;,
\label{Poissonbracket}
\end{equation}
is generalized to~\cite{Nambu:1973qe} 
\begin{equation}
\frac{\rd}{\rd t}\,F = \{F,H_1,H_2\}\;,   
\label{NambuCM}
\end{equation}
where the Nambu bracket on the right-hand side is defined as~\cite{Nambu:1973qe}
\begin{equation}
\{A,B,C\} = \sum_{ijk}\varepsilon_{ijk}\,
\dfrac{\partial A}{\partial q_i}
\dfrac{\partial B}{\partial q_j}
\dfrac{\partial C}{\partial q_k}
\;.
\label{NambuBracket}
\end{equation}
The Nambu bracket satisfies the so-called Fundamental Identity (the analog of the Jacobi identity for the Poisson bracket)~\cite{Nambu:1973qe, Takhtajan:1993vr}
\begin{eqnarray}
\{\{A,B,C\},D,E\} = 
\{\{A,D,E\},B,C\}
+ \{A,\{B,D,E\},C\} + \{A,B,\{C,D,E\}\},
\quad
\label{NambuFundamental}
\end{eqnarray}
which implies the Leibniz rule for the operator $\{*,H_1,H_2\}$,
justifying its identification with the time derivative.
{{The quantum version of Nambu's dynamics is given formally by}}
\begin{equation}
i\dfrac{\rd}{\rd t}\,F = [F,H_1,H_2]\;,
\label{NambuQM}
\end{equation}
where $F$, $H_1$, and $H_2$, are all to be properly defined objects in what follows (ultimately to be identified as cubic matrices),
and $[A,B,C]$ is the triple Nambu commutator (also called ``termutator'' in~\cite{Bhatta:2026wuz}) of such objects, defined as
\begin{equation}
[A,B,C] = ABC + BCA + CAB - ACB - BAC - CBA\;.
\end{equation}
The triple Nambu commutator maintains the antisymmetry of the Nambu bracket so that if any two entries are the same, it is automatically zero, ensuring that the two Hamiltonians are conserved since $[H_1,H_1,H_2]=[H_2,H_1,H_2]=0$.
(There exist obvious generalizations of the triple Nambu equation of motion as well as the Nambu triple bracket, etc for the case of $n$-tuple Nambu bracket, and Nambu's equations of motion, with $n{-}1$ Hamiltonians.)

Takhtajan introduced the action~\cite{Takhtajan:1993vr} for Nambu's generalized dynamics (for the case of triple Nambu bracket, but also immediately generalizable for the $n$-tuple bracket)
\be
S_N = \int_{C} (p\, \rd q_1 \wedge \rd q_2 
                - H_1\,\rd H_2 \wedge \rd t),
\ee
where $C$ denotes a
2-chain (a worldsheet swept out by a loop, i.e., a closed string) 
in the generalized phase space with a local {\em\/triple\/} of coordinates, $(q_1,q_2,p)$,
which suggests a connection between Nambu's dynamics and strings. Nambu's equations of motion follow from the least action principle by setting the variation of this action to zero, $\delta S_N =0$.
In the canonical example of classical Nambu dynamics represented by the asymmetric top
$H_1 = \frac{L_1^2}{2 I_1} + \frac{L_2^2}{2 I_2} + \frac{L_3^2}{2 I_3}$ and $H_2 = L_1^2 + L_2^2 + L_3^2$.
Of course the asymmetric top can be represented by the canonical Hamilton dynamics, based on the Poisson bracket.
Thus one might argue that the quantization of the Nambu bracket is just the usual quantization of very special (superintegrable) classical systems~\cite{Curtright:2002hm}.
Even though the usual deformation quantization of the classical Nambu bracket does not work, there is one example of formal deformation quantization of the Nambu bracket based on the so-called Zariski quantization~\cite{Dito:1996xr}. We will comment on the Zariski quantization in the next section devoted to string theory where, in the non-perturbative, M-theoretic formulation (related to membrane theory), one encounters the quantum Nambu bracket.

Apart from important formal issues there are major physics puzzles concerning Nambu's quantum dynamics:
What is the general physics meaning of the two Hamiltonians? Are there any distinguishing empirical effects associated with Nambu's quantum dynamics? How is Nambu's quantum theory related to the observed canonical quantum theory? Is quantum Nambu theory a consistent generalization of quantum theory?
Recently~\cite{Bhatta:2026wuz} it has been uncovered that Nambu's quantum theory leads to triple interference that is forbidden in canonical quantum theory, and that this effect is completely separate from observed canonical quantum effects. One can imagine that only upon extra (for example, gravitational) decoherence mechanism associated with purely cubic (and thus, beyond-quantum) matrix degrees of freedom can one observe a relation between Nambu's quantum theory and canonical quantum mechanics. The meaning of the cubic (rank-3) matrices that are required for the description of triple interference~\cite{Bhatta:2026wuz} and the physical meaning of two Hamiltonians might become apparent once one understands that in the context 
of gravity quantum measurement is radically different: that is, what is measured back-reacts, because information gathered in any measurement procedure is physical, it does have an energy content, and energy as a gravitational charge necessarily back-reacts~\cite{Berglund:2025qyo, Berglund:2025yoy}. This is why in the context of quantum gravity one needs new observables and new measurement procedures, and also why one Hamiltonian could be taken in the canonical sense and the other Hamiltonian as a suitably defined ``gravitational deformation'' of the original Hamiltonian of the system. This procedure generalizes to all orders for hypercubic observables and higher order back-reaction in measurement. We will elaborate on these points in concluding sections of the paper. Right now we will continue with the discussion of the relation between string theory and Nambu's dynamics.

\subsection{String theory, Membranes and M-theory}
String theory (originally proposed as a theory of strong interactions~\cite{Rickles:2014fha, Hubsch:2026The}) represents an approach to the problem of quantum gravity which naturally incorporates perturbative quantum gravity coupled to Standard Model-like gauge fields and chiral fermions~\cite{Polchinski:1998rq, Polchinski:1998rr, rBZ-StrTh}. Historically, the action for the string was given in terms of the area of its worldsheet (the Nambu--Goto action~\cite{Polchinski:1998rq}), which can be rewritten to explicitly involve the Poisson bracket
\be
S_{NG} = - \int \rd^{2} \sigma \sqrt{\tfrac12 \{X^{\mu},
X^{\nu} \}^{2}}.
\ee
but due to its non-polynomial nature (which can be made polynomial, but non-linear, of a Poisson--Yang--Mills type, using a Lagrange multiplier), the Nambu--Goto action was replaced in the 
{\em\/classically equivalent\/}
standard formulation of (bosonic) string theory using the Polyakov path integral~\cite{Polchinski:1998rq} (with $h_{ab}$(X) representing gravitational field living in classical spacetime with zero beta function equations for the running of the worldsheet ``coupling'' $h_{ab}$, required by conformal invariance, implying  Euclidean Einstein's GR equations~\cite{Polchinski:1998rq})
\begin{equation}
 \sum_{g} g_s^{2g-2} \int \frac{\rD X\, \rD\gamma}{\Vol{(\mathrm{Diff}\times\mathrm{Weyl})}}\,e^{-\int \sqrt{\gamma}\,\rd\tau\,\rd\sigma~\gamma^{\alpha\beta} (\pa_\alpha X^a)(\pa_\beta X^b) h_{ab}(X)  + \ldots}
\end{equation}
where the $X$-fields map from the worldsheet to the target spacetime of the string, $\gamma$ is the worldsheet metric, and $g_s$ is the string coupling while $g$ denotes the genus of the string worldsheet. The formulation is perturbative and involves an explicit sum over topologies as well as the quantization of two-dimensional gravity. (In the first quantized formalism, quantum field theory can be thought of as $0+1$ quantum gravity, and string theory as $1+1$ quantum gravity.)
The crucial point is that the worldsheet action of the string is renormalizable and the theory is conformally invariant (unlike other p-brane actions that appear as solitonic, collective, or bound states in string theory~\cite{Polchinski:1998rq, Polchinski:1998rr, rBZ-StrTh}). It is conformal invariance that is crucial in relating the beta function of the background fields to their target actions, most famously, the Einstein--Hilbert action for gravity~\cite{rF79a, rF79b, Braaten:1985is}). Note that conceptually string theory is defined as an S-matrix theory, harking back to its origins~\cite{Polchinski:1998rq, Polchinski:1998rr}. It is also naturally related to effective field theory of matter coupled to gravity, as follows from the requirement of conformal invariance which reproduces target space equations (and the associated effective action) for both matter and gravitational degrees of freedom in a background spacetime~\cite{Polchinski:1998rq, Polchinski:1998rr}. Upon compactification of higher dimensions or an effective (topological, geometrical or dynamical) isolation of a 3+1-dimensional sub-spacetime, one gets, in principle, an effective field theory description for Standard Model like matter coupled to Einsteinian gravity, in the observed four dimensional spacetime~\cite{Polchinski:1998rq, Polchinski:1998rr, rBZ-StrTh}; for a hierarchical sketch of the variety of possibilities, see Fig.~1 in Ref.~\cite{Berglund:2022qsb}. The observables in all so-obtained effective theory models descend from the stringy S-matrix observables. (One can even argue for uniqueness of string theory, based on S-matrix considerations~\cite{Cheung:2025tbr}. However, we will see that a deeper understanding of string theory reveals a surprising connection with quantum foundations and the appearance of completely new observables, which go beyond the canonical S-matrix that inherently correlates with the standard Born rule of quantum theory).

Note that 
there exists a plethora 
of possibilities for string compactifications 
(jointly dubbed the ``landscape'' or ``swampland'' of string 
models~\cite{Polchinski:2006gy}) with an, in principle, unknown measure on this ``space of models.'' 
String theory offers no guidance for computing such a measure, except for eliminating those effective field theory descriptions that are inconsistent from the point of view of string theory, for example, by being anomalous --- on the worldsheet or in the target space. In general, one faces the problems of predictability in the space of 
models (and especially, the prediction of the observed Standard Model of visible matter that lives in an accelerating universe with a tiny cosmological constant/vacuum energy),
and one has to resort to some general principles that are not inherent to string theory, for example, the general statistical principles, such as observer bias (also known as the ``anthropic principle'')~\cite{Polchinski:2006gy}. 
One of the 
aims
of this paper is to offer a new view on these foundational questions in string theory. These stem from purely stingy features of string theory that 
simply do not exist in
quantum field theory, such as T-duality or modular invariance, which also might imply a generalization of the uncertainty principle, at least, for stringy probes~\cite{Polchinski:1998rq, Chang:2011jj}. In particular, T-duality (together with modular invariance) will be our guide in the metastring formulation of string theory which will lead us to the foundations of quantum theory and the relevance of Nambu dynamics.

One of the main claims of this paper is that string theory stems from quantum foundations and that it represents a quantum theory that makes its own quantum geometry dynamical. In particular Nambu's dynamics and the Nambu bracket will appear in the non-perturbative description of such a formulation, suggesting both new observables, not captured by the perturbative formulation of string theory, as well as new dramatic effects associated with intrinsic triple interference of massive quantum objects. We note that foundations of quantum theory did appear previously in efforts to understand the question, ``What is string theory?''~\cite{Polchinski:1994mb}. For non-critical strings (that is, strings propagating through spacetime of non-critical dimension, in which the Liouville field theory appears, stemming from two-dimensional gravity) seen as a sum over random surfaces, there is a connection between the above Polyakov path integral (including the sum over topologies) to a matrix model. The matrix model appears after the worldsheet of the string is understood to emerge in the large $N$ limit of that matrix model, which can be associated to the dual triangulation of the worldsheet, seen as a random surface~\cite{Ginsparg:1993is, DiFrancesco:1993cyw, Polchinski:1994mb}.
The matrix model is of the kind that reminds one of matrix quantum mechanics, plus a certain conditions for criticality necessary for the achievement of the continuum limit for the worldsheet~\cite{Polchinski:1994mb}. However, the observables that appear in this picture are still canonical quantum observables based on the Born rule, such as the S-matrix.

There are natural superstring extensions of the Polyakov path integral which we will not write~\cite{Polchinski:1998rq, Polchinski:1998rr}, but we will comment on the origin and role of supersymmetry (especially in the context of quantum structure of spacetime) in what follows. The five original superstring theories can all be embedded into an M-theory~\cite{Polchinski:1998rr} whose conjectured light-cone formulation called M(atrix)-theory~\cite{Banks:1996vh} can be then related to Nambu's dynamics.
One simple way to realize this is that at strong coupling the string thickens (the thickness being proportional to the string coupling) and that one ends with a membrane with the following Polyakov %action
integral over histories
\begin{equation}
    \int \frac{\rD X\, \rD\gamma}{\Vol(\mathrm{Diff})}\,
    e^{-\int\sqrt{\gamma}\, \rd\tau\,\rd\sigma_1\,\rd\sigma_2~
    \gamma^{i j} (\pa_i X^A)(\pa_j X^B) h_{AB}(X)  + \ldots}
\end{equation}
where $X$-fields now map from the world-volume of the membrane into the target space, and $\gamma$ is the metric on the world volume.
The membrane, unlike the string, is not conformally invariant, so its path-integral measure reflects this fact.
Also, and unlike the worldsheet action of string theory, the world-volume action of the membrane is not renormalizable.
Furthermore, supersymmetry is crucial for the theory of membranes because of stability (bosonic membranes are unstable)~\cite{deWit:1988wri}. The similar reasoning is given for the bosonic string~\cite{Polchinski:1998rq, Polchinski:1998rr}, but we will revisit this in the section on the metastring.

\subsection{Membranes, M-theory and Matrix Models}
Even though the theory of membranes seems to have deep problems, there does exist a profound connection between the membrane theory and a non-perturbative formulation of string theory, called M theory, in particular, in the light-cone frame, via a matrix model formulation called M(atrix) theory, or the BFSS matrix model~\cite{deWit:1988wri, Banks:1996vh}. Likewise, the covariant version of the membrane can be directly related to Nambu's triple commutator, opening a possibility for the formulation of M-theory in terms of a quantized Nambu bracket. In the following we follow the old reviews~\cite{Jejjala:2007rn} and~\cite{Taylor:2001vb}.

Given the fact that conformal invariance is lacking in the theory of membranes, one can only partially gauge-fix the world-volume metric,
$\gamma_{00}=-\frac{4}{N^2}\textrm{det}
  \big[(\pa_i X^A)(\pa_j X^B) h_{AB}(X)\big]$, 
and
$\gamma_{0b}=0$,
where $N$ is a constant that will get a new meaning in the matrix regularization of the membrane.
For $\tau =\textit{const.}$, one may use the Poisson bracket (here defined over $\sigma^1$ and $\sigma^2$: 
$\{f,g\}\define\epsilon^{ab}(\pa_a f)(\pa_b g)$),
to rewrite the membrane action (in the flat spacetime background) as~\cite{Jejjala:2007rn},
\begin{equation}
S=\frac{N T}{4}\int \rd^3\xi \Big(\dot{X}^{\mu}\dot{X}_{\mu}-\frac{2}{N^2}\{X^{\mu},X^{\nu}\}\{X_{\mu},X_{\nu}\}\Big).
\end{equation}
The variation of the membrane action yields the following equations of motion
$\ddot{X}^{\mu}=\frac{4}{N^2}\{\{X^{\mu},X^{\nu}\},X_{\nu}\}$,
including the constraint equations
$\dot{X}^{\mu}\dot{X}_{\mu}=-\frac{2}{N^2}\{X^{\mu},X^{\nu}\}\{X_{\mu},X_{\nu}\}$,
as well as
$\dot{X}^{\mu}(\pa_aX_{\mu})=0 \Rightarrow \{\dot{X}^{\mu},X_{\mu}\}=0$.
The light-cone coordinates are used in order to address the issue of quantization,
$X^{\pm}=\frac{1}{\sqrt{2}}(X^0\pm X^{D-1})$,
where in the light-cone gauge,
$X^+=\tau$,
so that the constraint equations can be explicitly solved
$\dot{X}^-=\frac{1}{2}\dot{X}^i\dot{X}_i+\frac{1}{N^2}\{X^i,X^j\}\{X_i,X_j\}$, and $\partial_aX^-=\dot{X}^i(\pa_aX^-)$,
where the indices $i,j$ run from $1$ to $D{-}2$.
In the light-cone gauge the Hamiltonian reads as follows
\begin{equation}
H=\frac{N T}{4}\int \rd^2\sigma \Big( \dot{X}^i\dot{X}_i+\frac{2}{N^2}\{X^i,X^j\}\{X_i,X_j\}\Big),\qquad i,j=1,\dots,D{-}2,
\label{mh}
\end{equation}
with only one remaining constraint,
$\{\dot{X}^i,X_i\}=0$.
The non-linearity of the equations of motion makes the membrane quantization difficult and so one turns to the matrix regularization scheme~\cite{deWit:1988wri}.
In the matrix regularization, we map functions on the membrane's surface to $N\times N$ matrices.
For simplicity, one assumes the membrane surface to be the sphere, $S^2$, so the membrane surface at any fixed time may be described as a unit two-sphere with an associated $SO(3)$-invariant symplectic form.
The Cartesian coordinates of the sphere are denoted as $q_1,q_2,q_3$, where $\sum_i q_i^2=1$, and the symplectic structure is encoded in the Poisson bracket,
$\{q_a,q_b\}=\epsilon_{abc}q_c$.
Given the fact that the Lie algebras of $SU(2)$ and $SO(3)$ are identical,
the algebra of $SU(2)$ may be expressed in terms of the Lie bracket,
$[\mathbf{J}_a,\mathbf{J}_b]=i\epsilon_{abc}\mathbf{J}_c$.
The coordinate functions on the membrane are related to matrices that are an $N$-dimensional representation of the generators of $SU(2)$
$q_a\rightarrow\frac{2}{N}\mathbf{J}_a$,
and all functions on the sphere can be expanded in terms of spherical harmonics
$f(q_1,q_2,q_3)=\sum_{l,m}a_{lm}Y_{lm}(q_1,q_2,q_3)$,
which are themselves functions of the coordinates on the sphere
$Y_{lm}(q_1,q_2,q_3)=\sum b^{(lm)}_{a_1\dots a_l} %q_{a}\dots q_{l}$.
q_{a_1}\dots q_{a_l}$.
For the unit sphere the coefficients 
$b^{(lm)}_{a_1\dots a_l}$ 
are symmetric and traceless and also for spherical harmonics for $l<N$ we have
$Y_{lm}(q_1,q_2,q_3)\rightarrow\mathbf{Y}_{lm}=\big(\frac{2}{N}\big)^{l}\sum b^{(lm)}_{a_1\dots a_l} 
\mathbf{J}_{a_1}\dots \mathbf{J}_{a_l}$.
Thus, the following matrix correspondence is valid for an arbitrary function,
$f(q_1,q_2,q_3)\rightarrow \mathbf{F}=\sum_{l<N,m}a_{lm}\mathbf{Y}_{lm}$.
In the same vein there is a correspondence between the Poisson bracket and the commutator (the Lie bracket)
\begin{equation}
\{f,g\}\rightarrow -\frac{iN}{2}[\mathbf{F},\mathbf{G}], \qquad 
\frac{1}{4\pi}\int \rd^2\sigma\, f\rightarrow\frac{1}{N} \Tr\,\mathbf{F}.
\end{equation}
This procedure yields the following regularized Hamiltonian~\cite{deWit:1988wri, Taylor:1999qk, Jejjala:2007rn} (and references therein)
\begin{equation}
H=\frac{1}{2\pi \ell^3_{\rm Pl}}\Tr\Big(\frac{1}{2}\dot{\mathbf{X}}^i\dot{\mathbf{X}}_i- \frac{1}{4}[\dot{\mathbf{X}}^i,\dot{\mathbf{X}}^j][\dot{\mathbf{X}}_i,\dot{\mathbf{X}}_j]\Big).
\end{equation}
The equations of motion read as
$\ddot{\mathbf{X}}^i+[[\mathbf{X}^i,\mathbf{X}^j],\mathbf{X}_j]=0$
and the constraint
$[\dot{\mathbf{X}}^i,\mathbf{X}_i]=0.$
This procedure leads to a quantum theory (matrix model) whose fundamental degrees of freedom are $N\times N$ matrices with $U(N)$ symmetry, and even though we have assumed the membrane to have the topology of the two-sphere, this matrix regularization procedure can be generalized to membranes of arbitrary topology~\cite{deWit:1988wri}.

Let us briefly comment on the case of a supersymmetric membrane~\cite{Bergshoeff:1987cm, deWit:1988wri}, given the instability of a purely bosonic membrane.
There are eleven bosonic degrees of freedom corresponding to the embedding of the membrane, but 
the reparameterization invariance of the worldvolume gauges away three of these so that there are at the end eight bosonic degrees of freedom.
The fermions start out as thirty-two component Majorana (real) spinors, but the mass-shell condition and the $\kappa$-symmetry (fermionic diffeomorphisms) each halve the available degrees of freedom.
So, the resulting bosonic and fermionic degrees of freedom are organized as an $\CN{=}\,8$ supermultiplet of the three-dimensional worldvolume theory~\cite{Bergshoeff:1987cm, deWit:1988wri}.
This procedure yields the following Hamiltonian~\cite{deWit:1988wri},
\begin{equation}
H=\frac{N T}{4}\int \rd^2\sigma \Big( \dot{X}^i\dot{X}_i+\frac{2}{N^2}\{X^i,X^j\}\{X_i,X_j\} -\frac{2}{N}\theta^T\Gamma_i\{X^i,\theta\}\Big), \label{gogu}
\end{equation}
where $\theta$ denotes a sixteen component Majorana spinor of $\mathrm{Spin}(9)$, and $\Gamma_i$ are the $\Gamma$-matrices associated to the Clifford algebra.
In the light-cone gauge we have $\kappa$-symmetry gauge-fixed as follows: $\Gamma^+\theta=0$, so that the matrix regularization yields the following supersymmetric (super-Yang--Mills) matrix model Hamiltonian~\cite{deWit:1988wri}
\begin{equation}
H=\frac{1}{2\pi \ell^3_{\rm Pl}}\Tr\Big(\frac{1}{2}\dot{\mathbf{X}}^i\dot{\mathbf{X}}_i-
\frac{1}{4}[\dot{\mathbf{X}}^i,\dot{\mathbf{X}}^j][\dot{\mathbf{X}}_i,\dot{\mathbf{X}}_j]+
\frac{1}{2}\theta^T\Gamma_i[\mathbf{X}^i,\theta]\Big).
\label{bing}
\end{equation}

At this point we remark that, as noticed in~\cite{Banks:1996vh} (for a pedagogical review see~\cite{Taylor:2001vb}),
in the infinite momentum limit, M-theory is equivalent to the $N\rightarrow\infty$ limit of $N$ coincident D$0$-branes, given by $U(N)$ super-Yang--Mills theory.
This yields the Hamiltonian
\begin{equation}
H=\frac{R_{11}}{2}\Tr\Big(\dot{\mathbf{X}}^i\dot{\mathbf{X}}_i-
\frac{1}{2}[\dot{\mathbf{X}}^i,\dot{\mathbf{X}}^j][\dot{\mathbf{X}}_i,\dot{\mathbf{X}}_j]+
\theta^T\gamma_i[\mathbf{X}^i,\theta]\Big),
\end{equation}
which is equivalent to the Hamiltonian arrived at by light-cone quantization of the supermembrane, and so,
remarkably, D$0$-branes describe the M$2$-brane. %
Although the original conjecture~\cite{Banks:1996vh} was made in the $N\to\infty$ limit, it was later suggested that for finite-$N$, Matrix theory is equivalent to the discrete light-cone quantized (DLCQ) sector of M-theory --- see Ref.~\cite{Jejjala:2007rn} and references therein. In this picture, D$0$-branes are the partonic (``atomic'') degrees of freedom, out of which all other objects are made in the light-cone formulation of M-theory (in direct analogy with quarks and gluons in the context of QCD). Note also that stable flat spacetime emerges because of supersymmetry (as the moduli space of the supersymmetric Yang--Mills matrix model), which is an important conceptual lesson. Still, M(atrix) theory has open issues regarding the lack of explicit covariance, background independence and manifest T-duality (upon toroidal compactifications one obtains higher dimensional versions of the supersymmetric Yang--Mills theory, leading to the matrix string, etc, and ultimately to non-renormalizable theories). M(atrix) theory is holographic, but not as cleanly as is the case in the context of AdS/CFT correspondence~\cite{Maldacena:1997re, Gubser:1998bc} (also based on the properties of D-branes~\cite{Polchinski:1996fm, Polchinski:1998rq, Polchinski:1998rr}), and even in this canonical case, it is not clear whether holography is fundamental, or an artifact of a limited understanding of quantum gravity, and especially its observables and distinguishing empirical signals~\cite{Hubsch:2024agh, Berglund:2025qyo, Berglund:2025yoy, Hubsch:2026lot, Bhatta:2026wuz}.

\subsection{Covariant M-theory and the Nambu Bracket}
Let us now briefly address the question: Can M(atrix) theory be made covariant? It is here where we see
the relation between M-theory and Nambu's quantum bracket. (The intuitive idea~\cite{Awata:1999dz} is to discretize the world volume of the
membrane by a new algebraic object, such as the cubic matrix~\cite{Awata:1999dz}, and formulate a matrix model version of such general objects based on the Nambu bracket of cubic matrices, such that it reduces to M(atrix) theory with canonical square matrices in the light-cone gauge.) 
Such a question seems natural, owing to the fact that the classical membrane action (the Nambu--Goto action involving its world volume) can be written in terms of the classical Nambu bracket,
\be 
S_m = -\!\int \rd^{3} \xi \sqrt{{-}\tfrac16 \{X^{\mu},
X^{\nu},X^{\rho} \}^{2}},
\ee
(where $\mu = 0,1,...,10$, as appropriate for supermembranes, the bosonic membranes being inherently unstable), as well as the conceptual depiction in Figure~\ref{f:2-3}
\begin{figure}[h]
\centering\large
\begin{tabular}{ccl}
matrices/commutator &$\leftrightarrow$& 2D surface \\
$U(N)$ &$\leftrightarrow$& 2D APD \\[2mm]
&$\Downarrow$& \\[2mm]
cubes/triple Nambu bracket 
  &$\leftrightarrow$& 3D volume \\
unknown &$\leftrightarrow$& 3D VPD \\
\end{tabular}\vspace*{-3mm}
\caption{A schematic depiction of the extension to triple Nambu bracket dynamics: ``APD'' denotes ``area-preserving diffeomorphisms,'' matched by ``VPD'' for ``volume-preserving diffeomorphisms''}
\label{f:2-3}
\end{figure}
since the two indices of the square matrices in Matrix theory 
correspond to the discretized  Fourier indices on 
the membrane. (Note, that as in the case of the Nambu--Goto action for strings, this non-polynomial expression can be made polynomial, and of the Nambu--Yang--Mills type, by employing a Lagrange multiplier.)

So how does one make the BFSS model covariant~\cite{Minic:1999js}? It is here that the formal problem of the quantization of the Nambu bracket enters. Given the fact that the usual phase-space deformation quantization~\cite{Curtright:2011vw} based on the exponentiation of the classical Nambu bracket (the so-called Moyal bracket) fails, one has to try something different.
The only available formal quantization is the so-called Zariski quantization~\cite{Dito:1996xr}.
The basic idea of~\cite{Dito:1996xr} is to replace the usual product of functions in 
the classical Nambu bracket, which is Abelian, associative, distributive and 
respects the Leibnitz rule, by another (deformed) product with the same properties. 
Then the modified Nambu bracket defined by such a product will 
satisfy the same properties of the classical Nambu bracket. To implement this formal procedure one proceeds as follows (we follow our old summary~\cite{Minic:1999js}):

Let $N^{\rm irr}$ denote the set of real irreducible normalized polynomials of 
three variables $x^1, x^2, x^3$, and let $Z_0$ be a real vector space having a basis indexed by products of elements
of $N^{\rm irr}$. 
Denote the basis of $Z_0$ by $Z_{u_1...u_m}$ where $u_1...u_m$ are 
elements of $N^{\rm irr}$. The vector space $Z_0$ becomes an algebra by defining a
product, 
$Z_{u_1...u_m} \bullet Z_{v_1...u_n} 
 =Z_{u_1...u_m v_1...v_n}$,
 called the ``Zariski product''~\cite{Dito:1996xr}.
Next, let $Z_{\hbar}$ be the vector space of polynomials in 
${\hbar}$ (the deformation parameter) with coefficients in $Z_0$ and let
$\zeta(\sum_r \hbar^r u_r) = \sum_r \hbar^r Z_{u_r}.$
Then the deformed Zariski product is defined as
$Z_{u_1...u_m} \bullet_{\hbar} Z_{v_1...u_n} = 
\zeta((u_1,...u_m) \times_{\alpha} (v_1...v_n))$
where the operation $\times_{\alpha}$ for two 
irreducible polynomials $u_1$ and $v_1$ is given by the following expression
$u_1 \times_{\alpha} v_1 = \tfrac12 (u_1 * v_1 + v_1*u_1)$,
and where $*$ is the usual Moyal product (for a review of the Moyal product and the Wigner--Weyl--Moyal formulation of quantum theory in phase space consult~\cite{Curtright:2011vw}) with respect to two variables $x_1$ and
$x_2$.
In general, one has to fully symmetrize the $*$ product of irreducible
polynomials and also the deformed product should be extended from $Z_0$ to
$Z_{\hbar}$ by requiring that it annihilates the non-zero
powers
of the deformation parameter.
Next, let $E = Z_0[y^1, y^2, y^3]$ be the algebra of polynomials 
in three variables $y^1, y^2, y^3$ with coefficients in $Z_0$ and
let $A_0$ be a
subalgebra of $E$ generated by the following formal ``Taylor series''
in $E$
\be
J(Z_u) = Z_u + \sum_i y^i Z_{\partial_{i} u} +
\tfrac12 \sum_{i,j} y^i y^j Z_{\partial_{ij} u} +... ~\define~ 
\sum_n\tfrac1{n!} (\sum_i y^i \partial_i)^{n} (Z_{u})
\ee
where $\partial_i Z_u \define Z_{\partial_i u}$ and 
$u$ are normalized polynomials in three variables $x^1, x^2, x^3$ 
that can be uniquely factorized in $u_1,...,u_m$, and $\partial_i$ is a
derivative
with respect to $x^i$.
Now, define the following generalized derivative
$\Delta_{a}(J(Z_u)) \define J(Z_{\partial_a u})$,
where $a=1,2,3$, so that the classical Nambu bracket on $A_0$ can be defined as~\cite{Dito:1996xr}
\be
[A,B,C]_{\bullet} = \sum_{p} \epsilon(p)\,
(\Delta_{p_1}A) \bullet
(\Delta_{p_2}B) \bullet
(\Delta_{p_3}C),
\ee
for three elements $A,B,C$ from $A_0$.
It was shown in~\cite{Dito:1996xr} that there exists a deformation of this Nambu bracket
in the sense of deformation quantization.
So, let $E_{\hbar}$ be the algebra of polynomials of $\hbar$ (the deformation
parameter) with coefficients in $E$ defined above and
consider a subspace $A_{\hbar}$ of $E_{\hbar}$ consisting of polynomials in
$\hbar$ with coefficients in $A_0$.
The next step is to define a deformed product~\cite{Dito:1996xr}
\be
J(Z_u) \bullet_{\hbar} J(Z_v)
= Z_u \bullet_{\hbar} Z_v + 
\sum_i y^i (Z_{\pa_{i} u} \bullet_{\hbar} Z_v + 
Z_u \bullet_{\hbar} Z_{\partial_{i} v} )+
...
\ee
where the deformed Zariski product of basis elements is defined
as before.
Then one can formally define the quantum Nambu bracket on $A_{\hbar}$ as follows~\cite{Dito:1996xr}
\be
[A,B,C]_{\bullet_{\hbar}} = \sum_{p} \epsilon(p)\,
(\Delta_{p_1}A) \bullet_{\hbar}
(\Delta_{p_2}B) \bullet_{\hbar}
(\Delta_{p_3}C),
\ee
for three elements $A,B,C$ from $A_{\hbar}$.
It is shown in~\cite{Dito:1996xr} that this quantum Nambu bracket satisfies
all three fundamental properties: 
it is totally antisymmetric, it satisfies the 
derivation property and the Fundamental Identity.
Therefore, this construction formally solves the quantization problem.

Given this formal construction of the deformed Nambu bracket based on Zariski quantization, the following formal algebraic action was proposed for a covariant version of Matrix theory~\cite{Minic:1999js}
\be
S_M = - \int \rd^{3} \xi \sqrt{{-}\tfrac16 [X^{\mu},
X^{\nu},X^{\rho}]^{2}_{\bullet_{\sfrc1N}}},
\ee
or in terms of the gauge fixed metric
\begin{equation}
S_M=\frac{N T}{4}\int \rd^3\xi \Big(\dot{X}^{\mu}\dot{X}_{\mu}-\frac{2}{N^2} [X^{\mu},X^{\nu}]_{\bullet_{\sfrc1N}}\, [X_{\mu},X_{\nu}]_{\bullet_{\sfrc1N}}\Big),
\end{equation}
which in the light-cone gauge goes into the M(atrix) like action, provided one maps
$\int \rd^2~\sigma [X^{i},X^{j}]^2_{\bullet_{\sfrc1N}}$ into a matrix commutator $\int \rd t~ [\mathbf{X}^{i},\mathbf{X}^{j}]^2$.
(This would also imply that it is possible to map $\int \rd^{3} \xi \sqrt{{-}\tfrac16 [X^{\mu},
X^{\nu},X^{\rho}]^{2}_{\bullet_{\sfrc1N}}}$ into $\Tr [\hat{X}_c^{\mu},
\hat{X}_c^{\nu},\hat{X}_c^{\rho}]^2$, where $\hat{X}_c$s are cubic matrices to be precisely defined in what follows.)
Here $\mu = 0,1,...,10$ and $X^{\mu}$ are elements of
$A_{\sfrc1N}$, that is, they are 
polynomials in $\sfrc1N$ with coefficients in $A_0$ 
(recall that $A_0$ is the algebra of ``shifted'' polynomials of 
three variables, in this case,
$\xi_1, \xi_2, \xi_3$) and where $\sfrc1N$ is taken to be the deformation parameter of
the Zariski
quantization.
However, this formulation is extremely abstract and formal and it does not illuminate the main physics issues related to string theory and in particular, the question of 
the foundational physical principles of string theory, as well as its main, purely stringy, phenomenological implications.
That is why we turn to metastring formulation of string theory, which is T-duality covariant and intrinsically non-commutative and phase-space-like and which sheds light on the origin of string theory in quantum foundations.

%============================================================================
\section{From Strings and Metastrings to Nambu's Dynamics}
\label{s:Strings2Nambu}
One of the main points of this paper is that string theory is rooted in quantum foundations, as is suggested by the connection between the Polyakov path integral and matrix models, as well as the connection between M-theory and quantized Nambu dynamics. One of our main claims is that string theory is a quantum theory %that exists in 
with
a dynamical quantum geometry (called Born geometry~\cite{Freidel:2013zga, Freidel:2014qna}) in quantum (or modular) spacetime~\cite{Freidel:2015pka, Freidel:2016pls}, %that is 
--- woven by the string itself.
The observables of such a formulation are generalizations of the S-matrix observables, and they are captured by 
higher order algebraic objects (cubes and hypercubes) consistent with the requirements of Nambu's generalized dynamics~\cite{Bhatta:2026wuz}.
The measurement theory of this formulation of string theory is also corrected to include the non-Markovian nature of the measurement process as well as the back-reaction between measurement and the prepared states (new fundamental features characterizing quantum gravity)~\cite{Berglund:2025qyo, Berglund:2025yoy}, that can be encapsulated in cubic and hypercubic observables~\cite{Bhatta:2026wuz}. The results of quantum gravitational measurements are captured by quantum probability measures that are dynamical and contextual and that imply triple and higher level interference phenomena that simply do not exist in quantum mechanics and quantum field theory.
Finally, the generalized dynamical and contextual probability measures (that generalize the Born rule) possibly offer a new perspective on the problem of measures in the space of string compactifications.
In particular, we want to reinterpret the basic cubic vertex of string theory (the ``pants diagram'') as an indicator of intrinsic triple interference. Also, supersymmetry (SUSY) provides ``hardness'' of spacetime (and stability of flat spacetime) because of the fermionic nature of its ``supercoordinates,'' but it has to be hidden (``broken'') in some 
non-trivial way. We suggest that gravitization of quantum theory~\cite{Hubsch:2024agh, Hubsch:2026lot} could hide SUSY because of SUSY's sensitivity to global properties of ``quantum spacetime'' (which is intrinsically non-commutative and bosonic, but with fundamental non-commutativity between classical and dual spacetimes). The bosonic nature of spacetime and dual spacetime can be turned into fermionic properties of superspace as superstring is realized in terms of a compactified bosonic string on a suitable self-dual lattice~\cite{Casher:1985ra, Berglund:2021hbo}.

\subsection{Hints of Stringy Geometry}
The 1+1-dimensional worldsheet foundation of all fully interactive string theory models is unique among general classes of quantum field theories to have infinitely many conserved quantities~\cite{DiVecchia:1987jb}. Their existence is compatible with (in fact, indispensable in) the Nambu cubic and hypercubic extensions of the standard quantum-mechanical dynamics, such as discussed below, regarding~\eqref{NambuQM}.

The underlying geometry of all string theory models stems from the one spatial extension of strings, which evidently admits dynamical modes that travel in two opposite directions. These left- and right-moving modes are completely independent and decoupled in closed strings, but are correlated by the boundary conditions in open strings. The configuration space of strings is thus an orientation-generated double cover of the free loop-space of the ``target space(time),'' $\mathscr{X}$, through which the strings propagate, branched over the locus where the image of the string shrinks to a point so the two orientations appear indistinguishable --- which is $\mathscr{X}$ itself~\cite[Chapter~0]{rBeast2}. Indeed, target-spacetime physics is routinely constructed from the tensor product of Hilbert spaces built separately (and independently) from the left- and the right-moving modes~\cite{Casher:1985ra, Berglund:2021hbo} --- which then perforce include two separate (chiral boson) 0-modes, $x_L$ and $x_R$. Routinely, $x\define(x_L{+}x_R)$ is identified with the (classical background) 
$\mathscr{X}$-coordinates~\cite{Polchinski:1998rq}, while the complementary $\tilde{x}\define(x_L{-}x_R)$ is typically ignored (except for the 0-modes in ``twisted sectors'' in Landau--Ginzburg orbifolds~\cite{Vafa:1989xc, Intriligator:1990ua} which are indispensable in mirror symmetry~\cite{Berglund:1991pp}; see also Ref.~\cite{Berglund:2021hbo}). This built-in orientation-induced doubling in string theory eerily reminds of quantum systems inherently defined in two spatial dimensions --- such as the surface of a square-shaped membrane, albeit with additional restrictions, most notably the hallmark modular invariance of string theory, but also including those stemming from interactions with open strings and their correlation of left- and right-handed modes as mentioned above.

Further features of the underlying geometry of all string theory (most notably Born geometry) will be emerge clearly justified below and we defer a more complete analysis of the complete geometry for now.

\subsection{Metastrings and Modular Spacetime}
In order to address these foundational problems in string theory we start with the concept of metastring theory, a T-duality covariant, intrinsically non-commutative and chiral-phase-space-like formulation of string theory~\cite{Freidel:2013zga, Freidel:2014qna, Freidel:2015pka, Freidel:2017xsi, Freidel:2017wst, Freidel:2017nhg}.
To introduce a T-duality covariant formulation of string theory (the metastring) we follow the logic of~\cite{Freidel:2013zga} and start with the canonical Polyakov action in flat Minkowski spacetime
\be
\frac{1}{2\pi \alpha' } 
S_{P}(X) =\frac{1}{4\pi \alpha'}\int_{\Sigma} \eta_{\mu\nu} (*\rd X^{\mu} \wedge  \rd X^{\nu}),
\ee
where $*,d$ denote the Hodge dual and exterior derivative on the worldsheet, respectively.
We generally will refer to local coordinates on the worldsheet $\Sigma$ as $\sigma,\tau$, and we follow the traditional  interpretation of $X^\mu$ as local coordinates on a target space $M$, here with Minkowski metric. 
Now we rewrite the Polyakov action in the first order formalism
\be\label{PXactionFlat}
\hat{S} = \int_{\Sigma} \left( \bm{P}_\mu \wedge\rd X^\mu+ \frac{\alpha'}{2} \eta^{\mu\nu} (*\bm{P}_\mu \wedge \bm{P}_\nu) \right),
\ee
where $\bm{P}_{\mu}=P_{\mu}\rd\tau + Q_{\mu}\rd \sigma$ represents a one-form, and where the momentum carried by a closed  loop $C$ on the worldsheet $\Sigma$ is given by 
$p_{\mu}=\int_{C} \bm{P}_{\mu}$. 
By integrating out $\bm{P}_\mu$, we find
$*\bm{P}_\mu=\frac{1}{\alpha'} \eta_{\mu\nu} \rd X^\nu
$
which gives the Polyakov action plus a boundary term which is exactly the kernel of the Fourier transform,
$
{\hat{S}} =-\frac{1}{\alpha'} S_{P}(X)+ \int_{\pa\Sigma} X^\mu  \bm{P}_\mu,
$
that takes us from $X$ to $\bm{P}$.
By integrating out $X$ instead, we get
$\label{intXflat}
\rd \bm{P}_\mu=0,$ and so we can locally write 
 $
 \bm{P}_\mu= \rd Y_\mu.
 $
 We stress that this is true only locally, and in order to interpret it globally we must allow
 $Y_{\mu}$ to be multi-valued on $\Sigma$. In other words, $Y_{\mu}$ should carry, as compared to $X^{\mu}$, additional monodromies associated with each non-trivial cycle of $\Sigma$.
 Thus, the function $Y$ is only quasi-periodic with periods given by
 $
 \int_{C} \bm{P}_{\mu} = \int_{C}\rd Y_{\mu}=p_{\mu},
 $
  and the action involving $Y$ is essentially the Polyakov action
 $
 \hat{S} =-\alpha' S_{P}(Y).
 $
This action is weighted with $\alpha'$ instead of $1/\alpha'$ because $Y$ naturally lives in momentum space.
Therefore: starting from (\ref{PXactionFlat})
we see that integrating the one form  $\bm{P}$ leads to the spacetime Polyakov action, and that integrating $X$ leads to the momentum space Polyakov action. In order to get a phase space action that involves both coordinates and momenta a natural idea is to partially integrate out $\bm{P}$.  
Given the natural worldsheet space and time decomposition of $\bm P_{\mu}= P_{\mu} \rd \tau + Q_{\mu}\rd \sigma $ the action reads in conformal co-ordinates
$
\hat{S} = \int  P_{\mu} \partial_{\tau}{X}^{\mu} -  Q_{\mu} \partial_{\sigma}X^{\mu} + \frac{\alpha'}{2} (Q_{\mu}Q^{\mu}- P_{\mu}P^{\mu} ).
$
The equations of motion for $P,Q$ are simply  
$
\alpha' P = \partial_{\tau}{X}$,
$ \alpha' Q= \partial_{\sigma}X.
$
By integrating $Q$ only, we get the following action in the Hamiltonian form:
$
\hat{S}=\int P\cdot\pa_\tau X -\left(\frac{\alpha'}{2} P\cdot P +\frac1{2\alpha'} \partial_{\sigma}X\cdot \partial_{\sigma}X \right).
$
Let us introduce a momentum space coordinate $Y$ such that 
$\partial_{\sigma} Y= P$. Like $X$, this coordinate is not periodic; its quasi-period $ Y(2\pi) -Y(0) $  represents the string momentum. Using this coordinate the action becomes 
$
\int \partial_{\sigma}Y\cdot \partial_{\tau}{X} - \frac12 \left( \alpha' \partial_{\sigma}Y\cdot \partial_{\sigma}Y + \frac1{\alpha'} \partial_{\sigma}X \cdot \partial_{\sigma}X
\right).$
The main point is that in this action both $X$ and $Y$ are taken to be quasi-periodic.
The usual Polyakov formulation is recovered if one insists that $X$ is single-valued, and the usual T-duality formulation is recovered if one insists that quasi-periods of $X$ appear only along space-like directions and have only discrete values.
For simplicity we may assume that the background metric $G_{\mu\nu}$ and axion $B_{\mu\nu}$ are constant but arbitrary, and as suggested by the double field formalism~\cite{Siegel:1993th, Siegel:1993xq, Hull:1998vg, Hull:2004in, Hull:2009mi}, 
it is convenient to introduce  coordinates $\X^{A}\define (X^{\mu}/\sqrt{\alpha'} ,Y_{\mu}\sqrt{\alpha'} )^{T}$  on phase space $\cal P$,
together with two metrics: 
$H_{AB} \define   \left( \begin{matrix} (G- BG^{-1}B) & (BG^{-1}) \\ -(G^{-1}B) & G^{-1}   \end{matrix} \right)
$ and  $\eta_{AB} = \left( \begin{smallmatrix} 0 & 1 \\ 1 & 0  \end{smallmatrix} \right)$. Here $\eta$ is a  neutral metric and $H$ a generalized, and doubled, Lorentzian one. 
These data are related: in particular, if we define $J\define \eta^{-1}H$, then $J$  is an involutive  transformation preserving $\eta$,
that is, $J^{2}=1$, and  $\,\, J^{T}\eta J=\eta.$ 
In~\cite{Freidel:2013zga}  $(\eta,J)$ was called a {chiral } structure on $\cal P$, which is also endowed with a symplectic form $\omega$, (with generalized metric $H=\eta J$).
To see how the symplectic form $\omega$ must be added to $\eta$ and $H$ in order to form what we call Born geometry, one has to take into account the intrinsic non-commutativity of the doubled coordinates~\cite{Freidel:2017nhg, Freidel:2017wst} (where we are relabeling $Y$ as the dual coordinate $\tilde{X}$),
\be
[\hat{X}(\tau,\sigma_1),\hat{\tilde{X}}(\tau,\sigma_2)] = 2 \pi i \lambda^2,
\ee
which can be interpreted as coming from an integration constant obtained by integrating the canonical equal-time commutator 
 \be
 [\hat{X}(\tau,\sigma_1),\pa_\tau\hat{{X}}(\tau,\sigma_2)]=[\hat{X}(\tau,\sigma_1),\pa_{\sigma}\hat{\tilde{X}}(\tau,\sigma_2)] 
 = 2\pi i\hbar \alpha' \delta(\sigma_{12}),
 \ee 
 with respect to $\sigma_2$, (by noticing that the only value of this integration constant consistent with causality is $\pi\alpha'\hbar$, where $\alpha' \sim \lambda^2$).
 Note that $H = \omega I$, where $I^2=-1$ is the complex structure, well known from quantum mechanics. Also, $\eta = \omega K$, with $K^2=1$, so $K$ is the real structure.
Thus, using the already introduced notation from double field formalism, we have that 
\be
[\X^a,\X^b]=  i \lambda^2 \omega^{ab},
\ee
and this allows us, finally, to write the following metastring action
\begin{equation}
S^{\text{ch}}_{\text{str}}=
\int
\rd\tau \rd\sigma~
    \Big[(\pa_{\tau}{\X}^a) \big(\eta_{ab}(\X)+\omega_{ab}(\X)\big)
    -(\pa_\sigma\X^a)H_{ab}(\X)\Big] (\pa_\sigma\X^b), 
\end{equation}
where $\X^a\define (X^a/\lambda ,\tilde X_a/\lambda )^{T}$ 
are coordinates on a chiral phase-space like (doubled) target spacetime and, in general, $\eta, H,\omega$ are all dynamical. Note that, as it should be clear from the above discussion, $x^a, \tilde x_a$ stem from the left and right moving modes of the bosonic string,
\begin{equation}
    x^a\define x^a_L + x^a_R  \qquad\text{and}\qquad
    \Tilde{x}^a \define x^a_L - x^a_R.
\end{equation}
    In the context of a flat metastring we have 
 constant $\eta_{ab}$,  $H_{ab}$
 and $\omega_{ab}$ (zero $\omega_{ab}$ provides a connection to double field theory~\cite{Siegel:1993th, Siegel:1993xq, Hull:1998vg, Hull:2004in, Hull:2009mi})
\be\label{etaH0} 
	\eta_{ab} = \left( \begin{matrix} 0 & \delta \\ \delta^{T}& 0  \end{matrix} \right),\quad
H_{ab} =  \left( \begin{matrix} h & 0 \\ 0 &  h^{-1}  \end{matrix} \right),
\quad \omega_{ab} = \left( \begin{matrix} 0 & \delta \\ -\delta^{T}& 0  \end{matrix} \right).
\ee
The standard Polyakov action~\cite{Polchinski:1998rq} is obtained when setting $\omega_{ab}=0$ and integrating out the $\tilde x_a$ so that indeed
\begin{equation}
  S_P=\int \sqrt{\gamma}\,\rd\tau\rd\sigma~
  \gamma^{\alpha\beta} (\pa_\alpha X^a)(\pa_\beta X^b) 
  h_{ab}(X)  + \ldots
\end{equation}

The new feature in the metastring formulation of the bosonic string is intrinsic non-commutativity 
and so there is a new  Heisenberg algebra (the standard vertex operators now appear as representations of a Weyl--Heisenberg algebra, and there are no cocycles in the algebra of vertex operators in this formulation, as opposed to the textbook treatment~\cite{Polchinski:1998rq})
\begin{equation}
    [\X^a,\X^b]=  i \ell_s^2 \omega^{ab} \implies [X^a,\Tilde{X}^b]=i\delta^{ab} \ell_s^2,
\end{equation}
where we use the string length $\ell_s$ as $\lambda$.
Intuitively, for a non-local, string-like probe $\delta q \sim G_N \delta p \to \delta q \sim G_N \delta{\tilde{p}} \sim G_N \frac{1}{\delta{\tilde{q}}} \to \delta q \delta{\tilde{q}} \sim G_N \to [q, \tilde{q}] \sim i \ell_s^2$
as well as the standard commutators (with $[\Pi, \tilde{\Pi}]=0$)
\begin{equation}
    [\X^a,\Q_b]=  i \hbar \delta^a_b \implies [X^a,\Pi_b]=i\delta^a_b\hbar,\quad 
    [\Tilde{X}^a,\Tilde{\Pi}_b]=i\delta^a_b\hbar.
\end{equation}
Note, if Kalb--Ramond background $B_{ab}$ (stringy ``axion'') is constant, but non-zero, dual coordinates do not commute. (The Kalb--Ramond field originates from the symplectic structure $\omega$ of the metastring, upon an $O(d,d)$ transformation~\cite{Freidel:2017nhg, Freidel:2017wst}.) In general we have 
non-associativity~\cite{Gunaydin:2013nqa,Freidel:2017wst, Freidel:2017nhg}, which we think should be important in understanding the origin of the visible matter sector in string theory, following the prescient ideas of~\cite{Gunaydin:1973rs, Gunaydin:1974fb, Gursey:1996mj}.

What characterizes metastring theory as an intrinsically non-commutative formulation of string theory is the following fundamental commutator~\cite{Freidel:2016pls}
\begin{equation}
\X^{A}\define (x^a , \tilde{x}_a)^{T}, \quad [ \hat{\X}^a, \hat{\X}^b] = i \lambda^2 \omega^{AB} ,
\end{equation}
with $\frac{1}{2}\omega_{AB} d X^A d X^B = \frac{1}{\hbar} d \tilde{x}_{a} \wedge d x^{a}$, where 
$\omega_{AB} = - \omega_{BA}$ is the canonical symplectic form on phase space $\cal{P}$.
The Heisenberg group $H_{\cal{P}}$ is generated by Weyl operators~\cite{Freidel:2016pls}
($\K$ stands for the pair $(\tilde{k},k)$ and  $\omega(\K,\K')=k \cdot \tilde{k}' - \tilde{k}\cdot k'$)
\begin{equation}
W_{\K} \define e^{2 \pi i \omega( \K, \X)}.
\end{equation}
These form a central extension of the translation algebra and play the role of vertex operators in metastring theory~\cite{Freidel:2015pka, Freidel:2016pls}
\begin{equation}
W_{\K}W_{\K'} = e^{2 \pi i \omega( \K, \K' )} W_{\K + \K'}.
\end{equation}
In this formulation we also have that %states are sections of degree one 
\begin{equation}
W_{\K'} \Phi(\K) = e^{2 \pi i \omega( \K, \K' )} \Phi(\K + \K').
\end{equation}

The commuting subalgebra of the original non-commuting algebra $[ \hat{\X}^a, \hat{\X}^b] = i \lambda^2 \omega^{AB} $, can be completely described by self-dual lattices (endowed with the natural symplectic 
form ($\omega$) coming from the commutator bracket). These in turn represent a discretization of a covariant phase space
defined by $x$ and $\tilde{x}$ and when lifted to
the original non-commutative algebra, require extra data associated with the lift that is described by
a doubly orthogonal $O(d,d)$ metric $\eta$ (where $d$ denotes the spacetime dimension),
a symmetric counterpart of the antisymmetric $\omega$, associated with $Sp(2d)$ transformations.
In order to define the vacuum state on this self-dual lattice, we need a conformal structure
$O(2, 2(d{-}1))$ \cite{Freidel:2016pls}. This triplet is Born geometry of metastring theory
\cite{Freidel:2013zga, Freidel:2014qna}, which is defining property of modular spacetime, but it can be also associated with the modular representation~\cite{Aharonov:2005uc} of quantum theory
\cite{Freidel:2016pls}, and it naturally captures quantum non-locality that is consistent with causality, given
the quantum nature of the unitary operators and the fact that the triple intersection of $Sp(2d)$, $O(d,d)$
and $O(2, 2(d{-}1))$, gives the Lorentz group \cite{Freidel:2017xsi}.
In the context of quantum theory, the above vertex operators are associated with modular variables~\cite{Aharonov:2005uc}
which are used to explicitly demonstrate the non-local nature of quantum theory that is not in contradiction with classical
causality. Indeed, as claimed, the foundational concepts from quantum theory are made manifest in the metastring formulation of string theory, and in general, the Born geometry of metastring theory becomes dynamical, turning the metastring into
gravitized quantum theory with a dynamical Born rule~\cite{Freidel:2014qna}.

Coming back to the essential elements of Born geometry~\cite{Freidel:2013zga, Freidel:2014qna, Freidel:2015pka}, we remark that the symplectic structure $\omega$ found in
$\rd s^2_{\omega}= \frac{1}{2}\omega_{AB} \rd\X^A \rd\X^B$
is encoded in the canonical commutator between $x^{a}$ and $\tilde{x}_{a}$ (or the canonical
commutator between $q$ and $p$ in quantum theory).
Similarly, the generalized, quantum, doubled, metric $H$ comes from the Born rule in quantum theory
$\rd s^2_{H} = H_{AB} \rd\X^A \rd\X^B$.
For weak gravity, this metric reduces to the spacetime metric (where spacetime can be viewed as a specific polarization in modular spacetime), and so
due to gravity's extreme weakness, only classical spacetime metric is observed at low energies and long distances. That is why it makes sense that the perturbative string theory sees classical spacetime at large distances. 
Therefore, in this formulation the usual
dynamical spacetime metric is the low energy leftover of the quantum metric (in general, associated with the Born rule).
The polarization (or locality) metric $\eta$ encodes the distinction between spacetime-like and dual-spacetime-like
aspects of phase space (and in this sense it defines an analog of the ``causal'' structure in modular spacetime)
$\rd s^2_{\eta} = \eta_{AB} \rd\X^A \rd\X^B$.
This new metric captures the essence of relative (or observer dependent) locality~\cite{Amelino-Camelia:2011hjg, Amelino-Camelia:2011lvm}. When $\eta$ is constant we have absolute locality. Curving $\eta$ means ``gravitizing the quantum''~\cite{Freidel:2014qna}. (Note that in general relativity the spacetime polarization is fixed and the metric is dynamical, whereas in quantum theory polarization is dynamical and the metric, given by the Born rule, is fixed. Gravitized quantum theory (or background independent quantum theory~\cite{Minic:2003en, Minic:2003nx} makes both its polarization and its probability rule dynamical.)
In general all three elements of Born geometry, $\omega$, $\eta$ and $H$ are dynamical and curved (as well as contextual) in metastring theory. That fact implies that, in general, the Born rule is dynamical (and contextual) in string theory --- which is the essence of ``gravitized quantum theory'' with immediate experimental implications, such as intrinsic triple and higher order interference.
Last but not least, we emphasize that the Lorentz group (in $D$ spacetime dimensions) lies at the intersection of the symplectic, neutral and doubly orthogonal groups  \cite{Freidel:2016pls},
\be\label{orthinter}
 {\rm O}(1,d{-}1)={\rm Sp}(2d) \cap {\rm O}(d,d) \cap {\rm O}(2,2(d{-}1)),
\ee
which sheds new light on the origin of quantum theory through compatibility of the causal (Lorentz) structure and non-locality encoded in the discreteness of quantum spacetime (the fundamental length and time). This also captures the role of relative (observer-dependent) locality~\cite{Amelino-Camelia:2011hjg, Amelino-Camelia:2011lvm},
needed to resolve the apparent contradiction between discreteness of quantum spacetime and Lorentz symmetry. The observer-dependent spacetimes have to be in superposition in order to resolve this apparent contradiction. This is the origin of quantum interferences~\cite{Freidel:2016pls}. This implies that the results of measurement (necessarily classical, because the observer-dependent spacetimes are classical) are probabilistic, and that probability has to be consistent with equations of motion in spacetime. Its measure is maximally symmetric (like the symmetries of Minkowski spacetimes) and thus it leads to the Born rule. The overlap of the symplectic transformation from the equations of motion and the orthogonal transformations tied to the quadratic nature of the Born rule give the unitary evolution~\cite{Freidel:2016pls}. Note that from this viewpoint the classical spacetime polarization of perturbative string theory is just one of many, and in metastring theory, different observers see different spacetimes. In quantum string theory they are in superposition (thus implying quantum interferences and quantum probabilities determined by the Born rule and the canonical nature of perturbative quantum string theory) in order to reconcile the string fundamental length/time with Lorentz symmetry. In general, for dynamical spacetime, the whole quantum structure is ``gravitized,'' implying dynamical and contextual quantum probabilities and triple and higher order interferences~\cite{Hubsch:2024agh, Hubsch:2026lot}.

The zero modes of the metastring provide a model for the so-called metaparticle~\cite{Freidel:2018apz, Freidel:2019jor, Freidel:2021wpl}.
The metaparticle world-line action is (with the canonical particle action emerging in the limit $\mu \to 0$ and $\tilde{p} \to 0$) 
\begin{equation}
S_{\rm mp} = \int \rd t~ \Big[p_\mu\, \dot x^\mu +\tilde p^\mu\, \dot{\tilde x}_\mu +\lambda^2 p_\mu\, \dot{\tilde p}^\mu - 
\frac{N}{2}\big(p_\mu p^\mu +\tilde p_\mu \tilde p^\mu - m^2\big) +{\tilde{N}}\big(p_\mu \tilde p^\mu - \mu \big)\Big],
\end{equation}
where the central aspects of Born geometry can be found as follows: 
$\omega$ is in (``Berry-phase'') $p_\mu\, \dot{\tilde p}^\mu$, 
$\eta$ in the diffeomorphism constraint $p_\mu \tilde p^\mu = \mu$
and 
$H$ in the Hamiltonian constraint $p_\mu p^\mu +\tilde p_\mu \tilde p^\mu = m^2$.
Note also, that $\tilde{x}$, $[x, \tilde{x}] = i \lambda^2$, $\tilde{p}$, $[p, \tilde{p}] = 0$, and $[x, p] = i\hbar = [\tilde{x}, \tilde{p}]$.
The metaparticle can be understood as the quantum excitation a modular quantum field operator ${\hat{\phi}}_{\lambda} (x, \tilde{x})$ ({modular fields}, or {metafields}, which emerge from the intrinsic non-commutativity of string theory~\cite{Freidel:2017nhg, Freidel:2017wst, Freidel:2018apz, Freidel:2021wpl}),
\begin{equation}
{\hat{\phi}}(x) \to {\hat{\phi}}_{\lambda} (x, \tilde{x}),
\end{equation}
with $[x, \tilde{x}]\,{=}\,i \lambda^2$ implies that we are dealing with a covariant (and contextual) non-commutative field theory~\cite{Freidel:2017nhg, Freidel:2017wst, Minic:2020oho}, which is directly tied to the inherent non-commutativity of the metastring.
{Given this intrinsic non-commutativity of metastring theory we expect mixing of UV and IR physics, and a double Renormalization Group (RG) and thus, not a usual effective field theory (EFT) description~\cite{Grosse:2004yu}.}
Note also that classical spacetime label $x$ of canonical QFT represent a choice of (classical spacetime) polarization in modular (quantum) spacetime with 
a contextuality parameter $\lambda$.
(Averaging over dual spacetime $\tilde{x}$ implies, generically, a classical evolution from the unitary evolution in the spacetime basis.) For each field ${\phi}_{\lambda} (x, \tilde{x})$ there is a dual field ${\tilde{\phi}}_{\lambda} (x, \tilde{x})$ which does not commute with its partner, and upon the integration over the dual spacetime label, it is a candidate for the invisible matter sector (dual to the observed, visible, Standard Model sector)~\cite{Minic:2020oho, Berglund:2020qcu}.

A couple of more comments about the metastring are in order: as we have already noticed, given a pair $(H,\eta)$ one may consider the operator
$J\define \eta^{-1}H$ which by the consistency of string theory requires 
$J$ to be a chiral structure, that is, a real structure ($J^2=1$) compatible with $\eta$, implying that indeed $J$ is an $O(d,d)$ transformation (realizing generalized T-duality in target space). The symplectic $Sp(2d)$ $\omega$, the $O(d,d)$ $\eta$ and the $SO(2,2(d-1))$ $H$~\cite{Freidel:2013zga, Freidel:2014qna, Freidel:2015pka, Freidel:2015uug, Freidel:2016pls} unify the complex geometry of quantum theory with
the metrical geometry of general relativity and the symplectic geometry of canonical Hamiltonian dynamics. 
Note that in the metastring phase space-like formulation the local phase space coordinates $\X$ are {quasiperiodic}
$
\X^A (\sigma + 2\pi) = \X^A(\sigma) + \Delta^A,
$
where $\Delta^A$ is the corresponding quasiperiod (which either vanishes for the canonical Polyakov string or
is given by the winding number in the usual treatment of T-duality on compact spaces), and that is why the background of the metastring is not classical, but modular (quantum) spacetime. 
We stress that, given the fact that the worldsheet formulation of the metastring is chiral, even though the fields are
doubled, the central charges (left and right) are $c_L= c_R=D$ and we still have $D=26$ for criticality.
The metastring is not manifestly invariant under the worldsheet Lorentz transformations as it contains monodromies $\X^A (\sigma + 2\pi) = \X^A(\sigma) + \Delta^A$, but it flows to a Lorentz invariant fixed point under the double Renormalization Group which is familiar in non-commutative field theory~\cite{Grosse:2004yu}.
The familiar scattering amplitudes of canonical Polyakov's formulation of string theory~\cite{Polchinski:1998rq} have to be generalized in the context of the metastring to accommodate the fundamental T-covariance, intrinsic non-commutativity and the chiral phase-space-like structure~\cite{Freidel:2015pka}. This points to a more general set of observables that we will discuss below in the proposal for a non-perturbative formulation of the metastring, which also ties to our general insight about the need for gravitized quantum theory (such as the metastring) in quantum theory of gravity~\cite{Minic:2002pd, Minic:2003en, Minic:2004rj, Jejjala:2007rn, Freidel:2014qna, Berglund:2022skk, Hubsch:2024agh}.

The classical equations of motion of the metastring 
$
\partial_{\tau} \X^A - (J\, \partial_{\sigma} \X)^A=0,
$
imply the relation between momenta and monodromies $2 \pi P = J(\Delta)$.
There is, therefore, a {soldering} between worldsheet null coordinates $\sigma^{\pm} \define \sigma \pm \tau$ 
and the chiral target space structure
$
\partial_{\pm} \X^A - (P_{\pm} \X)^A=0,
$
where the chiral projector is defined as $2 P_{\pm} = (1\pm J)$.
This is what allows for decoupling of the left geometry from its right counterpart, which is somewhat reminiscent of twistor theory.
Also, the operator product expansion of the metastring vertex operators 
$
V_k = \epsilon_k e^{i \K\X},
$
(i.e. modular variables, or generalized Aharonov--Bohm phases~\cite{Aharonov:2005uc}) lead to the restriction of $\K$ on a double Lorentzian integral lattice $\Gamma$, that by modular invariance, must be
self-dual. These exist in $D\equiv2\pmod 8$, and are unique. Criticality gives a very unique lattice
$
\Gamma = \Pi_{1,25} \times \Pi_{1,25}.
$
This fact, in turn, leads to the large symmetry structure found in the
study of the monstrous moonshine~\cite{Moore:1993zc}.
As already noted, the metastring is chiral. This requires the introduction of a
preferred worldsheet time coordinate which is fundamentally
Lorentzian~\cite{Freidel:2015pka} and even though this might appear to be
inconsistent with modular invariance, it can be shown to be perfectly consistent.
Note that the metastring is bosonic, and that supersymmetry, and the superstring, can be viewed as emergent from this more fundamental formulation (see~\cite{Casher:1985ra}). 
In some sense, fermions can be viewed as ``defects or kinks'' in modular spacetime. The bosonic degrees of freedom associated with quanta of interactions can
be viewed as ``deformations'' of modular spacetime cells. 
The ``atoms ({\em\/ameres\/}) of spacetime''~\cite{Hubsch:2026The} (the fully compactified bosonic string) might have completely different statistics; given the appearance of the matrix model in what follows, the natural statistics for the atoms of spacetime, the actual modular spacetime cells, is the quantum distinguishable, or quantum Boltzmann statistics~\cite{Hubsch:2024agh, Hubsch:2026lot}, also found in the context of the non-commutative probability theory~\cite{Gopakumar:1994iq}. The fully compactified bosonic string (at strong coupling, as discussed in the next subsection), does not have any moduli and provides the perfect model for spacetime ``atoms'' which 
through the process of ``extensification''~\cite{Freidel:2015pka} grow spacetime through higher order correlations characteristic of triple and higher order coherent interference phenomena. This picture (involving dynamical and contextual generalizations of the Born rule for the probability measure) should be matched to the standard compactified picture from ten and eleven dimensions to the observed four dimensional spacetime, in the space of compactifications, characterized with measures that are responsible for the observed masses and interactions of (visible) elementary particles. Note that the fully compactified bosonic metastring does not have issues with closed time-like loops 
and causality (which was a concern in~\cite{Moore:1993zc}), because of the modular nature of time~\cite{Aharonov:2005uc}. Note also that the ``tachyon'' of perturbative bosonic string theory is precisely represented by the modular variable $e^{i \K\X}$, and it is not tied to the usual instability of the bosonic string action living in classical spacetime, because the metastring lives in a modular (quantum) spacetime~\cite{Freidel:2015pka}.

\subsection{Nonperturbative Metastring Theory and Nambu's Dynamics}
So far in our discussion, we have treated the worldsheet of the metastring itself as if a classical space.
However, there is no reason to not treat the worldsheet also as a modular space, if it is considered to be a ``quantum object.''
By making the world sheet a modular space (and thus na{\"\i}vely doubling it, by introducing $\tilde{\tau}$ and $\tilde{\sigma}$ in addition to $\tau$ and $\sigma$) and by Fourier-transforming dual world-sheet coordinates $\tilde{\tau}$ and $\tilde{\sigma}$  (as reviewed when we discussed membranes and M-theory), we end up with a formal matrix model realization of the metastring action~\cite{Berglund:2022qsb, Berglund:2023gur}
\be
S_{\text{meta}}=
\int
\rd\tau \rd\sigma~
    \Big[(\pa_{\tau}{\hat{\X}}^a) \big(\eta_{ab}(\hat{\X})+\omega_{ab}(\hat{\X})\big)
    -(\pa_\sigma\hat{\X}^a)H_{ab}(\hat{\X})\Big] (\pa_\sigma\hat{\X}^b). 
\ee
Also after using the dictionary for matrices $\pa_\sigma \cdot \define [\hat{\X}, \cdot]$, 
as well the already derived correspondence between Poisson brackets and commutators of matrices,
$\{f,g\}\rightarrow -\frac{iN}{2}[\mathbf{F},\mathbf{G}],$ as well as
$\frac{1}{4\pi}\int \rd^2\sigma\, f\rightarrow\frac{1}{N} \Tr\,\mathbf{F}$,
we obtain the following matrix model (Weyl-ordered) in one extra dimension ($26+1$ for the bosonic string)
\be
S^{\text{MM}}=
\int
\Big( (\pa_\tau \hat{{\X}}^i) [\hat{\X}^j, \hat{\X^k}] t'_{ijk}(\hat{\X})
    -[\hat{\X}^i, \hat{\X}^j] [\hat{\X}^k, \hat{\X}^l] H_{ijkl}(\hat{\X})\Big).
\ee
Here the ``triple'' $t'_{ijk}$ background combines the symplectic form $\omega$ and the bi-orthogonal metric $\eta$ of Born geometry. and the doubled metric $H$ has been absorbed in the $H_{ijkl}$ background.
(Note that there is a similarity here with Penrose's palatial twistor theory~\cite{Penrose:2015lla}. Also, the idea that the worldsheet of string theory should become ``non-commutative'' is quite old~\cite{Atick:1988si}, and it finds its explicit realization in the matrix model (MM) formulation of the metastring.)

Now we also use the already discussed insight that at strong coupling the string thickens into a membrane (the string coupling being the size of that thickness). So, one naturally ends up with matrix cubes that we have discussed in the case of membranes and M-theory! That can be seen after we remember that the Nambu bracket of three objects $A, B, C$ can be also written as
\be
[A,B,C] \define (\pa_{\tau} A)[B,C] + (\pa_{\tau} B)[C,A] + (\pa_{\tau} C)[A,B],
\ee
allowing us to obtain the final, fully covariant and ``cubic,'' form of the non-perturbative metastring (where $\X_c$ represents the cubic generalization~\cite{Bhatta:2026wuz} of 
square matrices that we will discussed in detail in the next section)
\begin{equation}
S^{\text{MM}}_{\text{GQ}}=
\int
\Big( [\hat{\X}_c^i, \hat{\X}_c^j, \hat{\X}_c^k] t_{ijk}(\hat{\X}_c)%\big)
    -[\hat{\X}_c^i, \hat{\X}_c^j] [\hat{\X}_c^k, \hat{\X}_c^l] H_{ijkl}(\hat{\X}_c)\Big),
 \label{e:SGQ}
\end{equation}
with $t_{ijk}$ totally antisymmetric, and $H_{ijkl} = H_{ik} H_{jl}$ doubling the metric of Born geometry.
Here we have used the label GQ to denote the gravitized quantum nature of this formulation of metastring theory.
This cubic 
matrix version contains information about dynamical (gravitized) geometry of quantum theory, which, as we shall see
in the next section is characterized 
with high-order interferences and
with a dynamical and contextual Born rule. (The cubic discretization of the membrane world volume might ultimately have something to do with the ``membrane-like'' $2+1$ dimensional nature of the general, background independent, ``holographic'' boundaries of the observed $3+1$ dimensional spacetime, but that is an open question.) The more interesting aspect is that the theory lives in a quantum spacetime, and that it has the structure of a gravitized quantum theory that is endowed with observables more general than the canonical S-matrix observables. The theory should also produce quantum measures different from the canonical Born measure, and thus the problem of measures on the space of string compactifications should be revisited from this new point of view. The theory does contain the Nambu bracket needed for the covariant formulation of M(atrix)/M-theory, and even though the theory is bosonic, given the fact that all superstring theories (different limits of M-theory) emerge from the bosonic string~\cite{Casher:1985ra}, a covariant formulation of M-theory should descend from this cubic matrix model formulation. Given the fact that T-duality is naturally incorporated in the metastring formulation, this particular shortcoming of the M(atrix) model formulation should be remedied as well. 
(The T-dual partner of M-theory is F-theory~\cite{Polchinski:1998rr}, whereby the above proposal naturally implies the appearance of hypercubical (rank-4) matrices. In fact, this might already be implied by the $H_{ijkl}(\hat{\X}_c)$ background in~\eqref{e:SGQ}, its $2{+}2$ split index structure indicative of the modular worldsheet as well as of the auxiliary torus needed for the geometric implementation of the $SL(2;\mathbb{Z})$ S-duality of the IIB-string.)

This non-perturbative formulation of the metastring should be compared to string field theory approaches to non-perturbative sting theory~\cite{Sen:2024nfd}. The non-perturbative metastring has the cubic (Nambu--Yang--Mills-like, or Chern--Simons-like part) and
a Yang--Mills part, both suggestive of open string field theory~\cite{Witten:1985cc} (see also~\cite{Strominger:1987ad}), appropriately ``gravitationally deformed.'' In particular, the pants diagram of closed string field theory should be understood as an evidence for the gravitized quantum theoretic nature of the metastring~\cite{Berglund:2022skk} from the point of view of the worldsheet. 
This fundamental (worldsheet) topology change is thus related to intrinsic triple interference. Similarly, the constituent picture found in the M(atrix) theory approach to a non-perturbative definition of string theory, is present in the above Nambu-like cubic matrix model formulation of the metastring. The elements of the cubic matrix are partons in this description, and only in the limit of 11 dimensional M-theory, with emergent supersymmetry, can they be related to D$0-$branes of M(atrix) theory. The issue of background invariance is tied to the gravitized nature of quantum metastring theory, in which the probability measure is dynamical and contextual (unlike the canonical one given by the Born rule, which violates the spirit of spacetime background independence), and its fundamental formulation, like that of dynamical spacetime geometry in general relativity, should be background invariant. Also, one should rethink the issue of background independence in the case of ``extensifying'' (or growing) quantum (modular) spacetime, and the fact that in the partonic picture (the entries of the cubic matrix) all backgrounds arise (emerge) from a different ``collective arrangement" of spacetime partons.

%============================================================================
\section{Nambu's dynamics, Metastrings and Gravitized Quantum Theory}
\label{s:Nambu2GQTh}
\subsection{Modular Spacetime and Foundations of Quantum Theory}

As noted in the previous section metastring vertex operators furnish a representation of the Heisenberg algebra, whose space of all commuting subalgebras (naturally related to commuting modular variables) is described by a self-dual lattice in phase space lifted to the Heisenberg algebra, with an associated Born geometry, that is, a modular spacetime.
Note that the same Born geometry, and the same modular variables appear in the foundations of quantum theory.
Thus metastring theory moves in a dynamical Born geometry (dynamical modular spacetime) characteristic of all quantum theories, except that the quantum (Born) geometry is dynamical in metastring theory. Thus, dynamical metastring theory (ideally, in a non-perturbative formulation) defines what we mean by ``gravitized quantum theory.''

Fundamentally, quantum theory originates from the compatibility of the fundamental length/time (defined by the string length/time) and Lorentz symmetry (implied by the compatibility of the basic elements of Born geometry), which implies observer-dependent spacetime and their superposition, and thus an inherent probabilistic interpretation with a maximally symmetric measure (given the maximal symmetry of flat spacetimes), and thus, the unique Born rule for the computation of quantum probabilities. The dual spacetime labels, when integrated over give an observed, classical, spacetime, with probabilistic events captured by the Born rule. In case of a dynamical spacetime (gravity) we have a
dynamical and contextual generalization of the Born rule --- that is a ``gravitization of quantum theory.''  This in turn implies intrinsic triple and higher order interferences, and new observables that needed to describe them.
We will see that such triple interferences are captured by the Nambu triple bracket structure found in the proposal for the non-perturbative definition of metastring theory.
So, string theory is only an S-matrix theory in classical spacetime. In general, it should lead to phenomena (such as intrinsic triple and higher order interference) not captured by the S-matrix formulation of quantum theory.

In the context of metastring theory the fundamental pants diagram should be an illustration of intrinsic triple interference of the theory associated with a cubic extension of the S-matrix variables --- S-cubes. Similarly the puzzling measure on the space of compactifications of string theory, should be (in the metastring context) related to a dynamical and contextual extension of the Born rule tied to triple and higher order interferences. Finally, supersymmetry should be fundamentally related the stability of the flat (but observer dependent) spacetime needed for the origin of canonical quantum interferences that are compatible with the fundamental length/time of string theory. But supersymmetry is hidden given the general curved nature of the ``gravitized quantum theory,'' as well as its M-theoretic origin in the bosonic string.
More directly, worldsheet $(0,1)$-supersymmetry suffices to eliminate vacuum instability (the tachyon), but cannot by itself induce supersymmetry in the target spacetime physics.

\subsection{Gravitized Quantum Theory and Intrinsic Triple Interference}
In this subsection we will demonstrate the relation between intrinsic triple interference, which is forbidden by the quadratic nature of the Born rule in quantum mechanics and quantum field theory, and Nambu's generalized dynamics.
First we say a few words about the general description of triple and higher order interference.

Let us stress that apart from the ``top-down'' rational for gravitized quantum theory provided by the metastring formulation of string theory, there is also a ``bottom-up" reason based on the geometry of quantum theory~\cite{Minic:2003en, Minic:2003nx, Minic:2004rj}.
{The geometry of quantum theory}~\cite{Ashtekar:1997ud}) is captured by maximally symmetric complex projective spaces $\mathbb{CP}^N$, with Fubini--Study (FS) metric, which is directly related to the maximally symmetric Fisher metric of information theory~\cite{Wootters:1981ki} as represented by the following expression
$\rd s_F^2 = \sum_i \frac{\rd p_i^2}{p_i}$, 
where $p_i$ are probabilities satisfying.
Setting $p_i \define \xi_i^2$ describes a sphere in a Euclidean space
$\rd s_F^2 = \sum_i \rd \xi_i^2,$ with 
$\sum_i \xi_i^2 = 1$.
If these $\xi_i$'s are the real and imaginary parts of the wavefunction, this defines an odd-dimensional sphere $S^{2N+1}$, which is always a $U(1)$ bundle over $\mathbb{CP}^N$, endowed with the compatible FS metric.
The compatibility of Fisher's metric with the symplectic form --- which governs the equations of motion --- gives the complex structure of the Schr\"{o}dinger equation.
The Schr\"{o}dinger equation can thus be seen as a geodesic equation on $\mathbb{CP}^n$ and the Born rule is the FS distance on $\mathbb{CP}^n$:
$\rd s_{FS}^2 (1,2) = 4(1\,{-}\,|\langle \psi_1| \psi_2 \rangle|^2)$.
Note that complex projective spaces are coset spaces: $\mathbb{CP}^n = U(n{+}1)/U(n) \times U(1)$. Here 
$U(n{+}1)$ represents unitary evolution,
$U(n)$ the general, non-Abelian, Berry phase and 
$U(1)$ --- the complex phase of the wave function.
One can understand also entanglement as a Segre embedding of the product of lower dimensional $\mathbb{CP}^{n}$s into a higher dimensional $\mathbb{CP}^{N}$. (In general, mixed states require the use of density operators/matrices, which are convex linear combinations of projectors, $\sum_ir_i|\varphi_i\rangle\langle\varphi_i|$ with $r_i\geqslant0$, over any suitable basis, $\{|\varphi_i\rangle\}$. The geometry of density matrices is then represented by the Bures metric --- a direct analogue of Fubini--Study/Fisher metric.)
Following Aharonov and Anandan~\cite{Aharonov:1987Pha} (see also~\cite{Minic:2003en, Minic:2003nx, Minic:2004rj}) one can also deduce the equation for a ``quantum clock'' where the probability measure (the Born rule) is related to the time interval measured by the quantum clock
$2 \hbar\,\rd s_{FS} = \Delta E\,\rd t$.
Here $\Delta E$ is the dispersion of energy,
and $\rd s_{FS}$ is the Fubini--Study metric on $\mathbb{CP}^n$
(or the statistical Fisher metric).
In the context of quantum spacetimes one should in general allow for topology change, in which case there is no unique timelike Killing vector, and thus $\Delta E$ is state dependent, and so the entire quantum geometry is state dependent, and therefore, dynamical and in general contextual, endowed with a dynamical and contextual Born rule. Also, for dispersion of energy that are of the order of the Planck energy $E_P$, one can only compensate the fully dynamical nature of clocks in general relativity with a fully dynamical nature of the probability measure, and so, one again, we are led to gravitized quantum theory, with a dynamical and contextual Born rule.

What is most important is the following observation: {Gravity violates the crucial underlying assumptions of Chentsov's theorem about the uniqueness of the Fisher metric}~\cite{Berglund:2025qyo, Berglund:2025yoy}, where the
Chentsov theorem states that: {The Fisher metric is the unique information metric under sufficient statistics and identical independent measurements}.  
Here by the word ``data'' we understand a set of permanent results from {independent identically distributed\/} (``i.i.d.'')\ measurements: however, in quantum gravity the data cannot be i.i.d., but instead the process of data gathering is intrinsically non-Markovian, as each successive recording of a data point affects the next measurement.
Also, sufficient statistics cannot be generated and passed between observers in quantum gravity due to different boundary conditions and a dynamical causal structure.
This implies a measurement procedure where the prepared state that leads to what is measured is back reacted on, by what is measured, and the whole measurement procedure is history dependent (non-Markovian, but in a form that is different from non-Markovian dynamics encountered in decoherence). This fundamental non-Markovian nature of measurement in quantum gravity, and thus in metastring theory, is also captured by higher order observables, cubic and hypercubic matrices that we have already mentioned in the context of the non-perturbative formulation of metastring theory, and that we will discuss in detail in what follows.

We follow Sorkin~\cite{Sorkin:1994dt} in thinking about triple and higher order interference phenomena.
In classical theory of probability, one has the
addition of probabilities
\be
P_{n}(A,B,C,\cdots) \,=\, P_{1}(A) + P_{1}(B) + P_{1}(C) + \cdots\;,
\ee
for any number of paths.
However, quantum mechanically, one has for two paths
$
P_{2}(A,B) = |\psi_A + \psi_B|^2 \vphantom{\Big|}$
or more explicitly
\be
{|\psi_A|^2} + 
{|\psi_B|^2} + 
{(\psi_A^*\psi_B^{\phantom{*}} + \psi_B^*\psi_A^{\phantom{*}})}
\define P_{1}(A) + P_{1}(B) + I_{2}(A,B),
\ee
where the last term
\be
I_{2}(A,B) = P_{2}(A,B)-P_{1}(A)-P_{1}(B),
\ee
is the ``interference'' of the two paths $A$ and $B$.
{Thus, non-vanishing double-path interference, 
$I_{2}(A,B)\neq 0$, distinguishes quantum from classical theory.}
In quantum theory, given its maximally symmetric geometry, the quadratic {Born rule}, fixed by this geometry, dictates that all the superimposed paths only interfere with each other 
in a pairwise manner.
For instance, for three paths we have
$P_{3}(A,B,C) = |\psi_A {+} \psi_B {+} \psi_C|^2$
\be
P_{2}(A,B) {+} P_{2}(B,C) {+} P_{2}(C,A) 
{-} P_{1}(A) {-} P_{1}(B) {-} P_{1}(C),
\label{eq:three-slit}
\ee
where only pairwise interferences between the pairs $(A,B)$, $(B,C)$, and $(C,A)$ appear.
It is clear from the above that in order for there to be a non-linear correction in an interference pattern the Born rule must be relaxed. Now, consider {a triple slit experiment}:
Since only pairwise interferences between the pairs $(A,B)$, $(B,C)$, and $(C,A)$ appear, it makes sense to define any deviation from this relation as the intrinsic
triple-path interference $I_{3}(A,B,C)$~\cite{Sorkin:1994dt}
\be
P_{3}(A,B,C)
-P_{2}(A,B)
-P_{2}(B,C)
-P_{2}(C,A)
+P_{1}(A)
+P_{1}(B)
+P_{1}(C).
\label{e:I3}
\ee
(This can be easily generalized for the case of $n$-paths.)
For both classical and quantum theory, this intrinsic triple-path interference is zero for any triplet of
paths. {Thus: experimental confirmation of $I_3=0$ would be a confirmation of the Born rule.}  
As already mentioned, rather weak bounds are currently placed on the parameter $\kappa$ that ``measures'' intrinsic triple interference ($\kappa \sim 10^{-3}$)
in photonic experiments, with 
\be
\kappa = \dfrac{\varepsilon}{\delta}, \quad \varepsilon =  I_3(A,B,C), \quad \delta  =  |I_2(A,B)| + |I_2(B,C)| + |I_2(C,A)|.
\ee

{{One of the drawbacks of above results is that triple interference ($I_3$) is not deduced based on a theory, because such a theory cannot be quantum mechanics (and of course, it cannot be classical either). So, the above limits are provided by the limitations of experimental probes of the Born rule. In order to give a consistent model of triple-interference~\cite{Bhatta:2026wuz}}} one has to invent new variables and new dynamics. Such a model can be formulated using ``density cubes'' (see~\cite{Dakic:2014ntj}),
in which the two-index density matrix $\rho_{ij}$ of canonical quantum mechanics (QM) is 
extended to the rank-three density cube
$\rho_{ijk}$,
where the subscripts are Hilbert space indices. As we will see in what follows, the dynamics of such cubes is given by Nambu's quantum dynamics.
Note that the inner product between two cubes $A_{ijk}$ and $B_{ijk}$ is defined as~\cite{Bhatta:2026wuz}
$(A,B) = \sum_{ijk} A_{ijk}^*B_{ijk},$
and the expectation value of observable $X_{ijk}$ 
for the density cube $\rho_{ijk}$ is given by
$\langle X\rangle_\rho = (\rho,X)$.
To ensure the reality of this expectation value, we %must 
demand both the density cube $\rho_{ijk}$ and
the observable $X_{ijk}$ to be ``Hermitian,'' 
where we define a cube $A_{ijk}$ to be Hermitian when
for any permutation $\pi$ of the three (arbitrary) indices it satisfies~\cite{Bhatta:2026wuz}
\begin{equation}
A_{\pi(ijk)} 
=
\bigg\{\begin{array}{@{}ll}
A_{ijk}   & \mbox{if $\pi$ is an even permutation,} \\
A_{ijk}^* & \mbox{if $\pi$ is an odd permutation.} 
\end{array}
\label{CubeHermiticity1}
\end{equation}
This guarantees that any transposition of the indices will complex conjugate the element so that
$(\rho,X)^* = (\rho,X)$,
and imposes the following conditions on the elements of a Hermitian cube~\cite{Bhatta:2026wuz}
\begin{equation}
\begin{array}{l}
A_{iii}\in\mathbb{R}\;,\quad
A_{iij} = A_{iji} = A_{jii} \in \mathbb{R}\;,\\
A_{ijk} = A_{jki} = A_{kij} = A_{ikj}^* = A_{jik}^* = A_{kji}^*\;.
\end{array}
\label{CubeHermiticity2}
\end{equation}
These are the rank-3 analogs of the conditions $A_{ii}\in\mathbb{R}$, $A_{ij}=A_{ji}^*$ for Hermitian matrices.
Define the ternary delta as~\cite{Bhatta:2026wuz}
\begin{equation}
\delta_{ijk} = \bigg\{\begin{array}{@{}rl}
1 & \mbox{when}~~i=j=k, \\
0 & \mbox{otherwise,}
\end{array}
\end{equation}
and the ``trace'' of a cube as
$\Tr\,A = (\delta,A) = \sum_{i}A_{iii}$.
The ``diagonal'' element $\rho_{iii}$ 
of the density cube is identified as the probability that the 
outcome associated with the
$i$th path is obtained, so they are normalized to
$\Tr\,\rho = (\delta,\rho) = \sum_{i}\rho_{iii} = 1\;$.
It is assumed that the basis of the Hilbert space is fixed since this relation is not invariant under basis transformations.
The elements with two distinct indices $\rho_{iij}$ and $\rho_{ijj}$ describe the interference between the $i$th and $j$th paths. These are specified by $N(N-1)$ real parameters, matching
the number for the density matrix.
Indeed, the diagonal and two-district-index elements of the 
density cube can be mapped to the elements of the density matrix $\rho_{ij}$ via
\begin{equation}
\rho_{iii}=\rho_{ii},\;\;
\rho_{iij}=\sqrt{\frac{2}{3}}\,\mathrm{Re}[\rho_{ij}],\;\;
\rho_{ijj}=\sqrt{\frac{2}{3}}\,\mathrm{Im}[\rho_{ij}],\;\;(i<j).
\label{MatrixCubeCorrespondence}
\end{equation}
The elements of the density cube with three distinct indices, $\rho_{ijk}$, are new degrees of freedom. 
They encode the triple-path interference of the $i$th, $j$th, and $k$th paths.

There are several requirements that the ternary product should satisfy~\cite{Bhatta:2026wuz}.
The first condition is that
it must be tri-linear in the three cubes
that constitute the product.
Second, the resulting termutator~\cite{Bhatta:2026wuz} (triple Nambu bracket) must be such that the right-hand side of Eq.~\eqref{NambuQM} is Hermitian 
so that the hermiticity of the density cube is maintained under time evolution.
Third, the termutator of Hermitian cubes must
be traceless so that the total probability %, Eq.~\eqref{TotalProb}, 
is also conserved.
Fourth, the fundamental identity must hold,
\begin{eqnarray}
[[A,B,C],D,E] = [[A,D,E],B,C] 
+ [A,[B,D,E],C] + [A,B,[C,D,E]]\;,
\label{FundamentalNambuCommutator}    
\end{eqnarray}
so that the operator $i[*,H_1,H_2]$ satisfies the Leibniz rule, just like the Nambu bracket.
The fifth and final requirement is the
ternary product to be ``associative'' in the sense that
\begin{equation}
(ABC)DE = A(BCD)E = AB(CDE) = A(DCB)E.
\label{Associativity}
\end{equation}
The following ternary product (see~\cite{Bhatta:2026wuz} and references therein) 
satisfies these properties:
\begin{equation}
(ABC)_{ijk} \;=\; \sum_\ell A_{ij\ell}B_{i \ell k}C_{\ell jk}\;,  
\label{KawamuraProduct}
\end{equation}
where repeated indices are not automatically summed, that is, $i,j,k$ are free rather than dummy indices.
Note that this is an obvious analogue of the product of two matrices. (An analogue of this works for hypercubes or rank-$n$ ``matrices.'')

Let $(E^{abc})_{ijk} \;=\; \delta^a_i\delta^b_j\delta^c_k,$ that is, 
$E^{abc}$ is an $N\times N \times N$ cube with the $abc$ element equal to $1$, and all other elements equal to $0$. Then, the space of all $N$-level Hermitian cubes is spanned by the 
following set of basis cubes~\cite{Bhatta:2026wuz}
\begin{alignat}9
E^a & =  E^{aaa}\,,\qquad
F^{ab} = \dfrac{1}{\sqrt{3}}\big(
E^{aab} + E^{aba} + E^{baa}
\big)\,,\\
J^{abc} & = \dfrac{1}{\sqrt{6}}
\big(E^{abc} + E^{bca} + E^{cab} + E^{acb} + E^{bac} + E^{cba}
\big) \,,\\
K^{abc} & = \dfrac{-i}{\sqrt{6}}
\big(E^{abc} + E^{bca} + E^{cab} - E^{acb} - E^{bac} - E^{cba}
\big)\,,
\end{alignat}
where differing superscripts have different values.
Note that $F^{ab}$ and $F^{ba}$ are different cubes, but
$J^{abc}$ and $K^{abc}$ are respectively symmetric and antisymmetric
under permutations of the superscripts.
For the basis $\{E^a,\,F^{ab},\,J^{abc},\,K^{abc}\}$,
the structure constants
\begin{equation}
i[A,B,C]\;=\;f_{ABCD}D,
\end{equation}
(of this ternary, or the Nambu--Lie algebra~\cite{Bhatta:2026wuz}) are found to be totally antisymmetric in the four indices, so we will only specify
the ones that cannot be obtained by a permutation of the four cubes.
They are~\cite{Bhatta:2026wuz}
\begin{eqnarray}
i[J^{abc},J^{abd},J^{acd}] & = &
i[J^{abc},K^{abd},K^{acd}] \; = \;
{-}\tfrac{1}{6}K^{bcd}\;,
\cr
i[J^{abc},J^{abd},K^{acd}] & = &
i[K^{abc},K^{abd},K^{acd}] \; = \; 
\tfrac{1}{6}J^{bcd},
\cr 
i[F^{ab},F^{ac},J^{abc}] 
& = & 
-\tfrac{1}{3}K^{abc}\;,
\label{TermutationRelations}
\end{eqnarray}
where $a$, $b$, $c$, and $d$ are all different.

Here are some explicit examples~\cite{Bhatta:2026wuz} of triple interference based on the fundamental Nambu equation for the 
density cube $\rho$ (which is the Nambu generalization of the usual von~Neumann equation for density matrix in quantum mechanics $\dfrac{\rd}{\rd t}\,\rho = i[\rho,H]$)
\begin{equation}
\dfrac{\rd}{\rd t}\,\rho = i[\rho,H_1,H_2]\;.
\label{NambuQMr}
\end{equation}
(Note that given the density cube $\rho$ we can define the cubic analogs of the von~Neumann entropy $-Tr(\rho \ln{\rho})$ as well as von Neuman's quantum information and entanglement entropy.)
For $N=3$, consider the subalgebra spanned by
$\{F^{12},F^{13},J^{123},K^{123}\}$, and let $H_1=F^{12}$ and $H_2=F^{13}$.
Let also
\begin{equation}
\begin{split}
\rho(t) =
\psi_1(t) J^{123} + \psi_2(t) K^{123}\;,
\end{split}
\end{equation}
where $\psi_{1,2}(t)$ are both real.
Then, Eq.~\eqref{NambuQMr} leads to~\cite{Bhatta:2026wuz}
\begin{eqnarray}
\dot{\psi}_1 = \frac{1}{3}\psi_2\;,\quad
\dot{\psi}_2 =  -\frac{1}{3}\psi_1\;,
\label{evolutioneq1}
\end{eqnarray}
\noindent
that is, $J^{123}$ and $K^{123}$ oscillate into each other.
If we are to avoid using $F^{ab}$ as the Hamiltonian,
then we must have $N\ge 4$ for ternary dynamics to exist.
For $N=4$, consider the subalgebra spanned by 
$\{J^{123},J^{124},J^{134},K^{234}\}$, and let 
$H_1 = J^{123}$ and  $H_2 = K^{234}$. Let also
\begin{equation}
\rho(t) = 
\psi_1(t)J^{124} + \psi_2(t)J^{134}\;,
\end{equation}
where $\psi_{1,2}(t)$ are real.
Then, Eq.~\eqref{NambuQMr} leads to~\cite{Bhatta:2026wuz}
\begin{eqnarray}
\dot{\psi}_{1} = \frac{1}{6}\psi_{2}\;,\quad
\dot{\psi}_{2} = -\frac{1}{6}\psi_{1}\;.
\label{evolutioneq4}
\end{eqnarray}
This describes the oscillation between
$J^{124}$ and $J^{134}$, which characterize the phenomenon of triple interference.
Thus, in this approach~\cite{Bhatta:2026wuz}, the Nambu triple commutator relations lead to
oscillations between pairs of triple-path interferences.
This intrinsic triple interference should be the smoking gun of a non-perturbative formulation of metastring theory in terms of the Nambu bracket as we now proceed to argue.

\subsection{The Metastring, Nambu's Dynamics and Intrinsic Triple Interference}
How would this reasoning apply to the metastring?
First we note that the metric $H_{ij}$ is going to satisfy an Einstein-like equation sourced by $t_{ijk}$.
The corresponding geodesic equation implied by Einstein's equations will be of the general form
\be
\frac{\rd\phi^i}{\rd s} + \Gamma^i_{jk}\frac{\rd\phi^j}{\rd s} \frac{\rd\phi^k}{\rd s} = t^{ijk}\, \phi_j \phi_k,
\ee
where $s$ is the appropriate evolution affine parameter associated with the line element of the geometry that solves Einstein's equations in the limit of classical spacetime.
This geodesic equation should be compared to
the so-called Nambu quantum theory~\cite{Minic:2002pd, Minic:2020zjb, Bhatta:2023biy}
\be
\frac{\rd\psi_a}{\rd\tau} = \Gamma_{abc}\, \psi_b \psi_c,
\label{e:deformDpsi}
\ee
where $\tau$ is the appropriate evolution parameter, which carries the same information (in the energy eigenbasis as the Nambu equation for the density cube, if we think of $\psi$ as being associated with ``pure states''). In the context of the canonical quantum theory Sch\"odinger's equation is the geodesic equation on the manifold of pure states, and it implies the von~Neumann equation for the density matrix for pure states. (Note also, that in the context of non-linear optics an effective triple interference phenomenon can be observed, using equations of the above form, by replacing $\psi$ with components of the electric field~\cite{Namdar:2021czo}.)
Thus, in the context of the proposed non-perturbative formulation of the metastring one should, in the cubic matrix picture, have a covariant analog of the Nambu equation for the density cube
\begin{equation}
\dfrac{\rd}{\rd s}\,\rho = i[\rho,H_1,H_2]\;.
\label{NambuQM-s}
\end{equation}
Here $\rd s^2 \define \Tr H_{ij}(\X_c) \rd\X_c^i \rd\X_c^j$ and the two Hamiltonians needed for Nambu's dynamics descend from the metastring Hamiltonian
$\partial_{\sigma} \X^A \partial_{\sigma} \X^B H_{AB}$, as well as the generator of dilatations, and in the matrix form read as
$H_1 \define [\hat{\X}_c^i, \hat{\X}_c^j] [\hat{\X}_c^k, \hat{\X}_c^l] H_{ijkl}(\X_c)$ (with $H_{ijkl}(\X_c) = H_{ik}(\X_c) H_{jl}(\X_c)$) and 
$H_2 \define [\hat{\X}_c^i, \hat{\X}_c^j, \hat{\X}_c^k] t_{ijk}(\X_c)$,
where $H_{ijkl}(\X_c)$ is determined by the solutions of the ``cubic matrix'' Einstein equation for $H_{ik}(\X)$, implied by the analog of the requirement of conformal invariance for metastring theory, and similarly for $t_{ijk} (\X_c)$.
These are the Hamiltonians in the interaction picture and $\rd s$ is the ``proper time'' in this Tomonaga--Schwinger version of Nambu's defining equation for the density cube. Note that the intrinsic non-commutativity of the metastring 
$[\X^a, \X^B] = i \lambda^2 \omega^{AB}$
would have to be generalized 
in the non-perturbative formulation to
\be
[\X_c^i, \X_c^j, \X_c^k] = i \lambda^3 t^{ijk} (\X_c),
\ee
with $t_{ijk}(\X_c)$ satisfying the equations of motion implied by the analog of the requirement of conformal invariance for metastring theory. 

Here by the analog of the requirement of conformal invariance, in the context of a matrix model formulation, we could require the vanishing of the beta function of the matrix analog of the Renormalization Group equations~\cite{Brezin:1992yc}. Even the defining Nambu equation for the density cube would have to be determined to be consistent with the geodesic equations for the quantum modular spacetime geometry of metastring theory. Also, integration of dual spacetime would provide a natural mechanism for the emergence of canonical quantum theory and its appropriate classical limit. We remark that the above Hamiltonians $H_1$ and $H_2$ represent both the pure open string sector as well as the coupling between the closed string (gravity) and open string sectors, which is physically appropriate for experiments involving massive quantum probes in which we claim new phenomena of intrinsic triple and higher order interference appear~\cite{Berglund:2023vrm}.

Indeed, if the $t^{ijk}$ background is given by $t^{ijk} = \epsilon^{ijkl} \X^l$
then we would end up with the ternary Nambu--Lie algebra considered in~\cite{Bhatta:2023biy} which, as reviewed above, combined with the equations of Nambu dynamics, implies the phenomenon of intrinsic triple interference!
Thus, the most important point is that the ``smoking gun'' for a non-perturbative formulation of string theory as gravitized quantum theory, as proposed in this work, could be found in intrinsic triple interference of massive quantum probes at low energy~\cite{Berglund:2023vrm}. In our previous work~\cite{Berglund:2023vrm, Hubsch:2024agh, Hubsch:2026lot} we have presented arguments regarding the scale of such low energy effects, to wit, the observed scale of vacuum (dark) energy, dynamically associated (from the metastring point of view) with the geometry of dual spacetime, which also plays role in the emergence of classical physics from its underlying quantum description, as is relevant for massive quantum probes, such as the proposed experiment involving falling nanoparticles and the non-linear Talbot effect~\cite{Berglund:2023vrm}.

We also note that the density cube can be related to the S-cube, in complete analogy with the formal relation between
the density matrix and the S-matrix~\cite{Hagston:1980xe}. This justifies the statement that non-perturbative string theory involves observables other than canonical S-matrices. Also, the closed string sector has an infinite number of interactions, which correspond to higher order interference phenomena, that can be all built, in principle, from the fundamental ``cubic interaction'' represented by the Nambu triple bracket. This procedure illustrates the claim that in the context of quantum gravity one needs new observables and new measurement procedures, and also why one Hamiltonian ($H_1$) could be taken in the canonical sense and the other Hamiltonian ($H_2$) as a suitably defined ``gravitational deformation'' of the original Hamiltonian of the system and why this procedure should be generalized to all orders for hypercubic observables and higher order back-reaction in quantum gravitational measurement.

We conclude with the following remark: In a recent paper~\cite{Pagani:2022pjq} it was noticed that the kernel of the von~Neumann equation for the density matrix (in the language of phase-space quantization a la Weyl--Wigner--Moyal~\cite{Curtright:2011vw}) can be written using the concept of a string in ``phase space'' (which is, apparently, unknown to the authors of~\cite{Pagani:2022pjq}, of the metastring form) with an effective action (the kernel being roughly the exponent of this action)
\be
S_{\text{ps-s}} \sim  \Big(\int_A \omega^{(2)} - \int_{\partial A} \rd H \Big),
\ee
where $H$ is the Hamiltonian of the system and $\omega^{(2)}$ is the natural two form on phase space, and $A$ denotes the area in phase space. It was also shown in~\cite{Pagani:2022pjq} that interference phenomena (say, the two-slit experiment involving electrons or photons) can be captured from this viewpoint. Given the fact that Nambu's dynamics involves the natural extension of the phase space with higher order forms (cubic etc) one can have an immediate extension for a membrane in ``phase space'' with an effective action
\be
S_{\text{ps-m}} \sim \Big(\int_V \omega^{(3)} - \int_{\partial V} \rd H_1 \wedge \rd H_2 \Big),
\ee
where $H_1$ and $H_2$ are the two Hamiltonian of Nambu's dynamics and $\omega^{(3)}$ is the natural three form on Nambu's phase space, and $V$ denotes the relevant volume of phase space. (This is immediately generalizable to higher orders.) It is natural to associate triple interference phenomena with an exponentiation of this last formula.

%============================================================================
\section{Conclusion: String Theory as Gravitized Quantum Theory}
\label{s:Coda}
In this paper we have proposed an answer to the old question: What is String Theory? in the form of the following
statement: String theory is Gravitized Quantum Theory of Modular Spacetime~\cite{Hubsch:2024agh, Hubsch:2026lot, Hubsch:2026The}.
We have argued that the non-perturbative version of such gravitized quantum theory is the cubic matrix model formulation of the metastring, a T-duality covariant and intrinsically non-commutative phase-space-like formulation of string theory
living in quantum (modular) spacetime.
Gravitational equations for quantum spacetime should emerge (as they do in canonical string theory) from the fixed point of a double RG which would control the dynamics (as a geodesic equation on quantum spacetime) of that gravitized quantum theory, with ``smoking gun'' phenomena of triple and higher interference associated with a dynamical and contextual Born rule. The theory has both spacetime and its dual, and thus the features required for addressing the question of quantum measurement, as well as the physics of the dual spacetime associated with dark matter and dark energy. The theory lives in modular spacetime and thus has the features needed for the correct evaluation of the vacuum energy as well as the parameters of the visible matter sector~\cite{Freidel:2022ryr, Freidel:2023ytq, Minic:2023oty, Berglund:2023gur, Hubsch:2024agh, Hubsch:2026lot}. As it is rooted in bosonic string theory in some limit it has the required visible degrees of freedom (without the requirement of supersymmetry, which is only an emergent feature, required for the stability of Minkowski spacetime patches). It may also contain the Standard Model degrees of freedom, after the reinterpretation of the usual $E_8$ theory (that arises in the heterotic string with chiral fermions, from the bosonic string reduced on a 16 dimensional lattice), as an octo-octonionic geometry with a real reduction to the real-octonionic geometry of the unique octonionic quantum theory, whose $SO(9)$ quantum phase does lead to the Standard Model gauge group when combined with the Poincare group in 4 spacetime dimensions, as pointed out by G\"unaydin and Gursey~\cite{Gunaydin:1973rs, Gunaydin:1974fb, Gursey:1996mj}. (The associated fermionic sector could emerge from the related complex-octonionic geometry and the $SO(10)$ quantum phase.) It also has the potential to describe the dual (invisible) matter sector, given the existence of dual fields that do not commute with standard local quantum fields of the Standard Model~\cite{Berglund:2020qcu, Minic:2020oho, Edmonds:2024qsj}. Similarly, curvature of dual spacetime can lead to a dynamical origin of dark energy~\cite{Berglund:2019ctg, Berglund:2020qcu, Hur:2025lqc}.

In the same framework one can address the observed cosmological constant~\cite{Freidel:2022ryr, Freidel:2023ytq} and masses of elementary particles~\cite{Minic:2023oty,Berglund:2023gur, Hubsch:2024agh, Hubsch:2026lot}. By utilizing the discretization of phase space
in terms of modular variables, where $N$ denotes the number of phase space cells, one can evaluate the value of vacuum energy as the volume of energy-momentum space as $N/l^4$, where $l^4$ is the spacetime volume (in four spacetime dimensions).
The number of phase space cells can be naturally related to entropy, because the modular variables  are commutative, but the number operator for modular position or momentum does not commute with the corresponding modular variable, and thus one can either be in a spacetime box or not (while not knowing in which one), and so the number of possibilities is $2^N$, thus giving $N$ an entropic interpretation. By equating this entropy with the gravitational one $N \sim l^2/\ell_P^2$ one obtains the scale for vacuum energy to be $\sqrt{l\,\ell_P}$, which for $l$ of the order of the observed cosmological size and $\ell_P$ the Planck length, gives the observed value of the cosmological constant.
This argument is particularly natural in the context of metastring theory which lives on modular spacetime~\cite{Freidel:2015pka, Freidel:2015uug, Freidel:2016pls}.

Other ``entropic'' scales can naturally emerge in the same approach~\cite{Minic:2023oty, Berglund:2023gur, Hubsch:2024agh, Hubsch:2026lot}:
the Bjorken--Zeldovich scale for quarks and charged leptons, and the neutrino scale (naturally related to the vacuum energy scale, as precisely needed to make contact with neutrino experiments).
These are needed for the explanation of other elementary particle masses, and the prediction of 
neutrino masses~\cite{Minic:2023oty, Berglund:2023gur, Hubsch:2024agh, Hubsch:2026lot}.
The Bjorken--Zeldovich scale, $l_{BZ}$, arises if gravitational entropy, which scales as $l^2/\ell_P^2$, is equated to the entropy of massive matter $l^3/l_{BZ}^3$, deduced to be $l_{BZ}^3 \sim l \ell_P^2$, and if $l$ is the size of the cosmological horizon, one finds that $l_{BZ} \sim 7\,\MeV$, around the scale of the lightest up and down quarks. 
Exactly the same mass-scale is also related by the seesaw relation to the Higgs mass-scale and the proton mass---which is the lightest (and so stable) fermionic bound-state of quarks~\cite{Hubsch:2024agh}!
Similarly, if one uses the masslessness of neutrinos in the classical Standard Model Lagrangian due to parity violation, then their entropy scales as the volume of spacetime $l^4/l_{\nu}^4$, where $l_{\nu}$ is the characteristic neutrino scale, which (if the neutrino entropy and the gravitational entropy $l^2/\ell_P^2$ are equated) is deduced to be $l_{\nu} \sim \sqrt{l\,\ell_P}$.
Identifying then $l$ with the cosmological horizon scale gives $l_{\nu} \sim \mu^{-1}$, that is the vacuum energy, or cosmological constant scale, compatible with the available evidence about the atmospheric and reactor neutrinos.
Thus, neutrinos could be viewed as ``portals'' between the Standard Model and the dark energy sector. 
Moreover, one can write the formulas for the neutrino masses (obeying the normal hierarchy) that involve the Planck and cosmological horizon scales, as well as the Standard Model scale~\cite{Minic:2023oty, Hubsch:2024agh, Hubsch:2026lot, Berglund:2023gur}.

We look forward with fascination to experimental responses to these predictions.

%============================================================================
\paragraph{Acknowledgments:}
Many thanks to P.~Berglund, N.~Bhatta, D.~Edmonds, J.~Erlich, L.~Freidel, A.~Geraci, P.~Huber, S.~Hur, N.~Ilic, V.~Jejjala, M.~Kavic, J.~Kowalski-Glikman, D.~Mattingly, H.~Minakata, R.~Pestes, D.~Stojkovic, T.~Takeuchi, C.~Tze as well
as K.~Nikolic and S.~Pajevic, and especially our dear late friend and collaborator Rob Leigh, for numerous insightful collaborations and illuminating discussions.
TH is grateful to the Department of Physics, University of Maryland, and the Physics Department of the University of Novi Sad, Serbia, for recurring hospitality and resources.
DM thanks Perimeter Institute in Waterloo, Ontario, Canada, for hospitality and support.
DM is supported by the U.S. Department of Energy (under contract DE-SC0020262).

\begingroup
\small\raggedright
%\bibliographystyle{utphys}
%\bibliography{../dSinST.bib}

\begin{thebibliography}{100}

\bibitem{Heisenberg:1925zz}
W.~Heisenberg, ``{A quantum-theoretical reinterpretation of kinematic and
  mechanical relations},'' \href{https://dx.doi.org/10.1007/BF01328377}{{\em Z.
  Phys.} {\bfseries 33} (1925) 879--893}.

\bibitem{Born:1925mph}
M.~Born and P.~Jordan, ``{Zur Quantenmechanik},''
  \href{https://dx.doi.org/10.1007/BF01328531}{{\em Z. Phys.} {\bfseries 34}
  no.~1, (1925) 858--888}.

\bibitem{Born:1926uzf}
M.~Born, W.~Heisenberg, and P.~Jordan, ``{Zur Quantenmechanik. II.},''
  \href{https://dx.doi.org/10.1007/BF01379806}{{\em Z. Phys.} {\bfseries 35}
  no.~8-9, (1926) 557--615}.

\bibitem{Dirac:1925jy}
P.~A.~M. Dirac, ``{The fundamental equations of quantum mechanics},''
  \href{https://dx.doi.org/10.1098/rspa.1925.0150}{{\em Proc. Roy. Soc. Lond.
  A} {\bfseries 109} (1925) 642--653}.

\bibitem{Born:1926yhp}
M.~Born, ``{Quantenmechanik der Sto{\ss}vorg{\"a}nge},''
  \href{https://dx.doi.org/10.1007/BF01397184}{{\em Z. Phys.} {\bfseries 38}
  no.~11-12, (1926) 803--827}.

\bibitem{Nambu:1973qe}
Y.~Nambu, ``{Generalized Hamiltonian dynamics},''
  \href{https://dx.doi.org/10.1103/PhysRevD.7.2405}{{\em Phys. Rev. D}
  {\bfseries 7} (1973) 2405--2412}.

\bibitem{Yoneya:2026Int}
T.~Yoneya, {\em Introduction to Nambu's Generalized Hamiltonian Dynamics},
  vol.~55 of {\em SpringerBriefs in Mathematical Physics}.
\newblock Springer, 2026.

\bibitem{Takhtajan:1993vr}
L.~Takhtajan, ``{On Foundation of the generalized Nambu mechanics (second
  version)},'' \href{https://dx.doi.org/10.1007/BF02103278}{{\em Commun. Math.
  Phys.} {\bfseries 160} (1994) 295--316},
  \href{https://arxiv.org/abs/hep-th/9301111}{{\ttfamily
  arXiv:hep-th/9301111}}.

\bibitem{Aaronson:2004ukm}
S.~Aaronson, ``{Is Quantum Mechanics An Island In Theoryspace?},''
  \href{https://arxiv.org/abs/quant-ph/0401062}{{\ttfamily
  arXiv:quant-ph/0401062}}.

\bibitem{Dowling:2002jjw}
J.~P. Dowling and G.~J. Milburn, ``{Quantum Technology: The Second Quantum
  Revolution},'' \href{https://dx.doi.org/10.1098/rsta.2003.1227}{{\em Phil.
  Trans. Roy. Soc. Lond. A} {\bfseries 361} no.~1809, (2003) 1655--1674},
  \href{https://arxiv.org/abs/quant-ph/0206091}{{\ttfamily
  arXiv:quant-ph/0206091}}.

\bibitem{Acin:2017gyd}
A.~Ac{\'\i}n {\em et~al.}, ``{The quantum technologies roadmap: a European
  community view},'' \href{https://dx.doi.org/10.1088/1367-2630/aad1ea}{{\em
  New J. Phys.} {\bfseries 20} no.~8, (2018) 080201},
  \href{https://arxiv.org/abs/1712.03773}{{\ttfamily arXiv:1712.03773
  [quant-ph]}}.

\bibitem{Zurek:2003zz}
W.~H. Zurek, ``{Decoherence, einselection, and the quantum origins of the
  classical},'' \href{https://dx.doi.org/10.1103/RevModPhys.75.715}{{\em Rev.
  Mod. Phys.} {\bfseries 75} (2003) 715--775},
  \href{https://arxiv.org/abs/quant-ph/0105127}{{\ttfamily
  arXiv:quant-ph/0105127}}.

\bibitem{Bose:2017nin}
S.~Bose, A.~Mazumdar, G.~W. Morley, H.~Ulbricht, M.~Toro\v{s}, M.~Paternostro,
  A.~Geraci, P.~Barker, M.~S. Kim, and G.~Milburn, ``{Spin Entanglement Witness
  for Quantum Gravity},''
  \href{https://dx.doi.org/10.1103/PhysRevLett.119.240401}{{\em Phys. Rev.
  Lett.} {\bfseries 119} no.~24, (2017) 240401},
  \href{https://arxiv.org/abs/1707.06050}{{\ttfamily arXiv:1707.06050
  [quant-ph]}}.

\bibitem{Marletto:2017kzi}
C.~Marletto and V.~Vedral, ``{Gravitationally-induced entanglement between two
  massive particles is sufficient evidence of quantum effects in gravity},''
  \href{https://dx.doi.org/10.1103/PhysRevLett.119.240402}{{\em Phys. Rev.
  Lett.} {\bfseries 119} no.~24, (2017) 240402},
  \href{https://arxiv.org/abs/1707.06036}{{\ttfamily arXiv:1707.06036
  [quant-ph]}}.

\bibitem{Penrose:2014nha}
R.~Penrose, ``{On the Gravitization of Quantum Mechanics 1: Quantum State
  Reduction},'' \href{https://dx.doi.org/10.1007/s10701-013-9770-0}{{\em Found.
  Phys.} {\bfseries 44} (2014) 557--575}.

\bibitem{Zurek:2020ukz}
K.~M. Zurek, ``{On vacuum fluctuations in quantum gravity and interferometer
  arm fluctuations},''
  \href{https://dx.doi.org/10.1016/j.physletb.2022.136910}{{\em Phys. Lett. B}
  {\bfseries 826} (2022) 136910},
  \href{https://arxiv.org/abs/2012.05870}{{\ttfamily arXiv:2012.05870
  [hep-th]}}.

\bibitem{Berglund:2023vrm}
P.~Berglund, A.~Geraci, T.~H{\"u}bsch, D.~Mattingly, and D.~Minic, ``Triple
  interference, non-linear {T}albot effect and gravitization of the quantum,''
  \href{https://dx.doi.org/10.1088/1361-6382/ace14a}{{\em Class. Q. Grav.}
  {\bfseries 40} no.~15, (2023) 155008 (24pp)},
  \href{https://arxiv.org/abs/2303.15645}{{\ttfamily arXiv:2303.15645
  [gr-qc]}}.

\bibitem{Weinberg:1989Tes}
S.~Weinberg, ``Testing quantum mechanics,''
  \href{https://dx.doi.org/https://doi.org/10.1016/0003-4916(89)90276-5}{{\em
  Annals of Physics} {\bfseries 194} no.~2, (1989) 336--386}.

\bibitem{Polchinski:1991Wei}
J.~Polchinski, ``Weinberg's nonlinear quantum mechanics and the
  einstein-podolsky-rosen paradox,''
  \href{https://dx.doi.org/10.1103/PhysRevLett.66.397}{{\em Phys. Rev. Lett.}
  {\bfseries 66} (Jan, 1991) 397--400}.

\bibitem{Mielnik:2000ej}
B.~Mielnik, ``{Comments on: ``Weinberg's Nonlinear Quantum Mechanics and
  Einstein-Podolsky-Rosen paradox'', by Joseph Polchinski},''
  \href{https://arxiv.org/abs/quant-ph/0012041}{{\ttfamily
  arXiv:quant-ph/0012041}}.

\bibitem{Minic:2003en}
D.~Minic and H.~C. Tze, ``{Background independent quantum mechanics and
  gravity},'' \href{https://dx.doi.org/10.1103/PhysRevD.68.061501}{{\em Phys.
  Rev. D} {\bfseries 68} (2003) 061501},
  \href{https://arxiv.org/abs/hep-th/0305193}{{\ttfamily
  arXiv:hep-th/0305193}}.

\bibitem{Minic:2002pd}
D.~Minic and H.~C. Tze, ``{Nambu quantum mechanics: A Nonlinear generalization
  of geometric quantum mechanics},''
  \href{https://dx.doi.org/10.1016/S0370-2693(02)01865-8}{{\em Phys. Lett. B}
  {\bfseries 536} (2002) 305--314},
  \href{https://arxiv.org/abs/hep-th/0202173}{{\ttfamily
  arXiv:hep-th/0202173}}.

\bibitem{Minic:2003nx}
D.~Minic and H.~C. Tze, ``{A general theory of quantum relativity},''
  \href{https://dx.doi.org/10.1016/j.physletb.2003.11.054}{{\em Phys. Lett. B}
  {\bfseries 581} (2004) 111--118},
  \href{https://arxiv.org/abs/hep-th/0309239}{{\ttfamily
  arXiv:hep-th/0309239}}.

\bibitem{Minic:2004rj}
D.~Minic and H.~C. Tze,
  \href{https://dx.doi.org/10.1142/9789812702340_0020}{``{What is quantum
  theory of gravity?},''} in {\em {3rd International Symposium on Quantum
  Theory and Symmetries}}, pp.~159--166.
\newblock 1, 2004.
\newblock \href{https://arxiv.org/abs/hep-th/0401028}{{\ttfamily
  arXiv:hep-th/0401028}}.

\bibitem{Jejjala:2007rn}
V.~Jejjala, M.~Kavic, and D.~Minic, ``{Time and M-theory},''
  \href{https://dx.doi.org/10.1142/S0217751X07036981}{{\em Int. J. Mod. Phys.
  A} {\bfseries 22} (2007) 3317--3405},
  \href{https://arxiv.org/abs/0706.2252}{{\ttfamily arXiv:0706.2252 [hep-th]}}.

\bibitem{Freidel:2014qna}
L.~Freidel, R.~G. Leigh, and D.~Minic, ``{Quantum Gravity, Dynamical Phase
  Space and String Theory},''
  \href{https://dx.doi.org/10.1142/S0218271814420061}{{\em Int. J. Mod. Phys.}
  {\bfseries D23} no.~12, (2014) 1442006},
\href{https://arxiv.org/abs/1405.3949}{{\ttfamily arXiv:1405.3949 [hep-th]}}.
%%CITATION = ARXIV:1405.3949;%%.

\bibitem{Minic:2020oho}
D.~Minic, ``{From quantum foundations of quantum field theory, string theory
  and quantum gravity to dark matter and dark energy},'' in {\em {10th
  MATHEMATICAL PHYSICS MEETING}: {School and Conference on Modern Mathematical
  Physics}}, pp.~183--218.
\newblock Institute of Physics, Belgrade, Serbia, 2020.
\newblock \href{https://arxiv.org/abs/2003.00318}{{\ttfamily arXiv:2003.00318
  [hep-th]}}.

\bibitem{Hubsch:2024agh}
T.~H{\"u}bsch and D.~Minic, ``Quantum gravity as gravitized quantum theory,''
  in {\em Proceedings of the 3rd Conference on Nolinearity (Sep. 4--8.\ 2023)},
  B.~Dragovich and {\v{Z}}.~{\v{C}}upi{\'c}, eds., pp.~1--72.
\newblock Serbian Academy of Nonlinear Sciences, Belgrade, Serbia, 2024.
\newblock \href{https://arxiv.org/abs/2407.06207}{{\ttfamily arXiv:2407.06207
  [hep-th]}}.

\bibitem{Rickles:2014fha}
D.~Rickles, \href{https://dx.doi.org/10.1007/978-3-642-45128-7}{{\em {A Brief
  History of String Theory. From Dual Models to M-Theory}}}.
\newblock The Frontiers Collection. Springer, Heidelberg, Germany, 2014.

\bibitem{Hubsch:2026The}
T.~H{\"u}bsch and D.~Minic, {\em The Weight of Quantum: Quantum Theory and the
  Structure of Physics}.
\newblock CRC Press, July, 2026.
\newblock
  \url{https://www.routledge.com/The-Weight-of-Quantum-Quantum-Theory-and-the-Structure-of-Physics/Hubsch-Minic/p/book/9781041191285}.

\bibitem{Berglund:2025qyo}
P.~Berglund, A.~Geraci, T.~Hubsch, D.~Mattingly, and D.~Minic, ``Information
  metrics and possible limitations of local information objectivity in quantum
  gravity,'' \href{https://dx.doi.org/10.1088/1361-6382/ae811a}{{\em Class. Q.
  Grav.} {\bfseries 43} no.~13, (2026) 135021 (9pp)},
  \href{https://arxiv.org/abs/2501.19269}{{\ttfamily arXiv:2501.19269
  [gr-qc]}}.

\bibitem{Berglund:2025yoy}
P.~Berglund, A.~Geraci, T.~Hubsch, D.~Mattingly, and D.~Minic, ``Physical
  limits on information metrics and quantum gravity as gravitized quantum
  theory,'' \href{https://arxiv.org/abs/2504.12925}{{\ttfamily arXiv:2504.12925
  [gr-qc]}}.

\bibitem{Sorkin:1994dt}
R.~D. Sorkin, ``{Quantum mechanics as quantum measure theory},''
  \href{https://dx.doi.org/10.1142/S021773239400294X}{{\em Mod. Phys. Lett. A}
  {\bfseries 9} (1994) 3119--3128},
  \href{https://arxiv.org/abs/gr-qc/9401003}{{\ttfamily arXiv:gr-qc/9401003}}.

\bibitem{Bhatta:2026wuz}
N.~Bhatta, D.~Minic, and T.~Takeuchi, ``{Dynamics of the density cube},''
  \href{https://dx.doi.org/10.1103/bn2b-jgll}{{\em APS Open Sci.} {\bfseries 1}
  (2026) L000165}, \href{https://arxiv.org/abs/2606.02421}{{\ttfamily
  arXiv:2606.02421 [quant-ph]}}.

\bibitem{Bergshoeff:2025haj}
E.~A. Bergshoeff, E.~Sezgin, and P.~K. Townsend, {\em {A brief history of
  supermembranes}}.
\newblock 3, 2025.
\newblock \href{https://arxiv.org/abs/2503.06234}{{\ttfamily arXiv:2503.06234
  [hep-th]}}.

\bibitem{Bergshoeff:1987cm}
E.~Bergshoeff, E.~Sezgin, and P.~K. Townsend, ``{Supermembranes and
  Eleven-Dimensional Supergravity},''
  \href{https://dx.doi.org/10.1016/0370-2693(87)91272-X}{{\em Phys. Lett. B}
  {\bfseries 189} (1987) 75--78}.

\bibitem{Bergshoeff:1988hw}
E.~Bergshoeff, E.~Sezgin, Y.~Tanii, and P.~K. Townsend, ``{Super $p^-$branes as
  Gauge Theories of Volume Preserving Diffeomorphisms},''
  \href{https://dx.doi.org/10.1016/0003-4916(90)90381-W}{{\em Annals Phys.}
  {\bfseries 199} (1990) 340}.

\bibitem{Polchinski:1998rq}
J.~Polchinski, \href{https://dx.doi.org/10.1017/CBO9780511816079}{{\em {String
  theory. Vol. 1: An introduction to the bosonic string}}}.
\newblock Cambridge Monographs on Mathematical Physics. Cambridge University
  Press, Cambridge, UK,
Dec., 2007.
\newblock
%%CITATION = INSPIRE-487240;%%.

\bibitem{Polchinski:1998rr}
J.~Polchinski, \href{https://dx.doi.org/10.1017/CBO9780511618123}{{\em {String
  theory. Vol. 2: Superstring theory and beyond}}}.
\newblock Cambridge Monographs on Mathematical Physics. Cambridge University
  Press, Cambridge, UK, Dec., 2007.

\bibitem{rBZ-StrTh}
B.~Zwiebach, {\em A First Course in String Theory}.
\newblock Cambridge University Press, 2004.

\bibitem{Cremmer:1978km}
E.~Cremmer, B.~Julia, and J.~Scherk, ``{Supergravity Theory in 11
  Dimensions},'' \href{https://dx.doi.org/10.1016/0370-2693(78)90894-8}{{\em
  Phys. Lett. B} {\bfseries 76} (1978) 409--412}.

\bibitem{deWit:1988wri}
B.~de~Wit, J.~Hoppe, and H.~Nicolai, ``{On the Quantum Mechanics of
  Supermembranes},''
  \href{https://dx.doi.org/10.1016/0550-3213(88)90116-2}{{\em Nucl. Phys. B}
  {\bfseries 305} (1988) 545}.

\bibitem{Banks:1996vh}
T.~Banks, W.~Fischler, S.~H. Shenker, and L.~Susskind, ``{M theory as a matrix
  model: A Conjecture},''
  \href{https://dx.doi.org/10.1103/PhysRevD.55.5112}{{\em Phys. Rev.}
  {\bfseries D55} (1997) 5112--5128},
\href{https://arxiv.org/abs/hep-th/9610043}{{\ttfamily arXiv:hep-th/9610043
  [hep-th]}}.
%%CITATION = HEP-TH/9610043;%%.

\bibitem{Ishibashi:1996xs}
N.~Ishibashi, H.~Kawai, Y.~Kitazawa, and A.~Tsuchiya, ``{A Large N reduced
  model as superstring},''
  \href{https://dx.doi.org/10.1016/S0550-3213(97)00290-3}{{\em Nucl. Phys.}
  {\bfseries B498} (1997) 467--491},
\href{https://arxiv.org/abs/hep-th/9612115}{{\ttfamily arXiv:hep-th/9612115
  [hep-th]}}.
%%CITATION = HEP-TH/9612115;%%.

\bibitem{Awata:1999dz}
H.~Awata, M.~Li, D.~Minic, and T.~Yoneya, ``{On the quantization of Nambu
  brackets},'' \href{https://dx.doi.org/10.1088/1126-6708/2001/02/013}{{\em
  JHEP} {\bfseries 02} (2001) 013},
  \href{https://arxiv.org/abs/hep-th/9906248}{{\ttfamily
  arXiv:hep-th/9906248}}.

\bibitem{Bagger:2006sk}
J.~Bagger and N.~Lambert, ``{Modeling Multiple M2's},''
  \href{https://dx.doi.org/10.1103/PhysRevD.75.045020}{{\em Phys. Rev. D}
  {\bfseries 75} (2007) 045020},
  \href{https://arxiv.org/abs/hep-th/0611108}{{\ttfamily
  arXiv:hep-th/0611108}}.

\bibitem{Bagger:2007jr}
J.~Bagger and N.~Lambert, ``{Gauge symmetry and supersymmetry of multiple
  M2-branes},'' \href{https://dx.doi.org/10.1103/PhysRevD.77.065008}{{\em Phys.
  Rev. D} {\bfseries 77} (2008) 065008},
  \href{https://arxiv.org/abs/0711.0955}{{\ttfamily arXiv:0711.0955 [hep-th]}}.

\bibitem{Gustavsson:2007vu}
A.~Gustavsson, ``{Algebraic structures on parallel M2-branes},''
  \href{https://dx.doi.org/10.1016/j.nuclphysb.2008.11.014}{{\em Nucl. Phys. B}
  {\bfseries 811} (2009) 66--76},
  \href{https://arxiv.org/abs/0709.1260}{{\ttfamily arXiv:0709.1260 [hep-th]}}.

\bibitem{Aharony:2008ug}
O.~Aharony, O.~Bergman, D.~L. Jafferis, and J.~Maldacena, ``{N=6 superconformal
  Chern-Simons-matter theories, M2-branes and their gravity duals},''
  \href{https://dx.doi.org/10.1088/1126-6708/2008/10/091}{{\em JHEP} {\bfseries
  10} (2008) 091}, \href{https://arxiv.org/abs/0806.1218}{{\ttfamily
  arXiv:0806.1218 [hep-th]}}.

\bibitem{Dito:1996xr}
G.~Dito, M.~Flato, D.~Sternheimer, and L.~Takhtajan, ``{Deformation
  quantization and Nambu mechanics},''
  \href{https://dx.doi.org/10.1007/BF02509794}{{\em Commun. Math. Phys.}
  {\bfseries 183} (1997) 1--22},
  \href{https://arxiv.org/abs/hep-th/9602016}{{\ttfamily
  arXiv:hep-th/9602016}}.

\bibitem{Sato:2012ti}
M.~Sato, ``{Zariski Quantization as Second Quantization},''
  \href{https://dx.doi.org/10.1103/PhysRevD.85.126012}{{\em Phys. Rev. D}
  {\bfseries 85} (2012) 126012},
  \href{https://arxiv.org/abs/1202.1466}{{\ttfamily arXiv:1202.1466 [hep-th]}}.

\bibitem{Minic:1999js}
D.~Minic, ``{M theory and deformation quantization},''
  \href{https://arxiv.org/abs/hep-th/9909022}{{\ttfamily
  arXiv:hep-th/9909022}}.

\bibitem{Curtright:2002hm}
T.~L. Curtright and C.~K. Zachos, ``{Nambu dynamics, deformation quantization,
  and superintegrability},'' in {\em {Workshop on Superintegrability in
  Classical and Quantum Systems}}.
\newblock 11, 2002.
\newblock \href{https://arxiv.org/abs/math-ph/0211021}{{\ttfamily
  arXiv:math-ph/0211021}}.

\bibitem{Zachos:2003md}
C.~K. Zachos and T.~L. Curtright, ``{Deformation quantization of Nambu
  mechanics},'' \href{https://dx.doi.org/10.1063/1.1594405}{{\em AIP Conf.
  Proc.} {\bfseries 672} no.~1, (2003) 183--196},
  \href{https://arxiv.org/abs/quant-ph/0302106}{{\ttfamily
  arXiv:quant-ph/0302106}}.

\bibitem{Curtright:2003fn}
T.~Curtright and C.~K. Zachos, ``{Quantizing Dirac and Nambu brackets},''
  \href{https://dx.doi.org/10.1063/1.1594404}{{\em AIP Conf. Proc.} {\bfseries
  672} no.~1, (2003) 165--182},
  \href{https://arxiv.org/abs/hep-th/0303088}{{\ttfamily
  arXiv:hep-th/0303088}}.

\bibitem{Curtright:2003wj}
T.~L. Curtright and C.~K. Zachos,
  \href{https://dx.doi.org/10.1142/9789812702340_0025}{``{Branes, strings, and
  odd quantum Nambu brackets},''} in {\em {3rd International Symposium on
  Quantum Theory and Symmetries}}, pp.~206--217.
\newblock 12, 2003.
\newblock \href{https://arxiv.org/abs/hep-th/0312048}{{\ttfamily
  arXiv:hep-th/0312048}}.

\bibitem{Zachos:2004zc}
C.~K. Zachos and T.~Curtright, ``{Branes, Quantum Nambu Brackets and the
  Hydrogen Atom},'' \href{https://dx.doi.org/10.1007/s10582-004-9807-x}{{\em
  Czech. J. Phys.} {\bfseries 54} no.~11, (2004) 1393--1398},
  \href{https://arxiv.org/abs/math-ph/0408012}{{\ttfamily
  arXiv:math-ph/0408012}}.

\bibitem{Curtright:2011vw}
T.~L. Curtright and C.~K. Zachos, ``{Quantum Mechanics in Phase Space},''
  \href{https://dx.doi.org/10.1142/S2251158X12000069}{{\em Asia Pac. Phys.
  Newslett.} {\bfseries 1} (2012) 37--46},
  \href{https://arxiv.org/abs/1104.5269}{{\ttfamily arXiv:1104.5269
  [physics.hist-ph]}}.

\bibitem{Berglund:2022skk}
P.~Berglund, T.~H\"ubsch, D.~Mattingly, and D.~Minic, ``{Gravitizing the
  Quantum},'' \href{https://dx.doi.org/10.1142/S021827182242024X}{{\em Int. J.
  Mod. Phys. D} {\bfseries 31} no.~14, (3, 2022) 2242024 (9 pages)},
  \href{https://arxiv.org/abs/2203.17137}{{\ttfamily arXiv:2203.17137
  [gr-qc]}}.

\bibitem{Minic:2020zjb}
D.~Minic, T.~Takeuchi, and C.~H. Tze, ``{Interference and Oscillation in Nambu
  Quantum Mechanics},'' \href{https://arxiv.org/abs/2012.06583}{{\ttfamily
  arXiv:2012.06583 [hep-ph]}}.

\bibitem{Huber:2021xpx}
P.~Huber, H.~Minakata, D.~Minic, R.~Pestes, and T.~Takeuchi, ``{Neutrino
  oscillations at JUNO, the Born rule, and Sorkin\textquoteright{}s triple path
  interference},'' \href{https://dx.doi.org/10.1103/PhysRevD.105.115013}{{\em
  Phys. Rev. D} {\bfseries 105} no.~11, (2022) 115013},
  \href{https://arxiv.org/abs/2105.14061}{{\ttfamily arXiv:2105.14061
  [hep-ph]}}.

\bibitem{Bhatta:2023biy}
N.~Bhatta, D.~Minic, and T.~Takeuchi, ``{Constraints on a generalization of
  geometric quantum mechanics from neutrino and B$^{0}$-$ \overline{B^0} $
  oscillations},'' \href{https://dx.doi.org/10.1007/JHEP02(2024)031}{{\em JHEP}
  {\bfseries 02} (2024) 031},
  \href{https://arxiv.org/abs/2310.07457}{{\ttfamily arXiv:2310.07457
  [hep-th]}}.

\bibitem{Freidel:2013zga}
L.~Freidel, R.~G. Leigh, and D.~Minic, ``{Born Reciprocity in String Theory and
  the Nature of Spacetime},''
  \href{https://dx.doi.org/10.1016/j.physletb.2014.01.067}{{\em Phys. Lett.}
  {\bfseries B730} (2014) 302--306},
\href{https://arxiv.org/abs/1307.7080}{{\ttfamily arXiv:1307.7080 [hep-th]}}.
%%CITATION = ARXIV:1307.7080;%%.

\bibitem{Freidel:2015pka}
L.~Freidel, R.~G. Leigh, and D.~Minic, ``{Metastring Theory and Modular
  Space-time},'' \href{https://dx.doi.org/10.1007/JHEP06(2015)006}{{\em J. High
  Energ. Phys.} {\bfseries 06} (2015) 006},
\href{https://arxiv.org/abs/1502.08005}{{\ttfamily arXiv:1502.08005 [hep-th]}}.
%%CITATION = ARXIV:1502.08005;%%.

\bibitem{Freidel:2015uug}
L.~Freidel, R.~G. Leigh, and D.~Minic, ``{Modular spacetime},''
\href{https://dx.doi.org/10.1142/S0218271815440289}{{\em Int. J. Mod. Phys.}
  {\bfseries D24} no.~12, (2015) 1544028}.
%%CITATION = IMPAE,D24,1544028;%%.

\bibitem{Freidel:2016pls}
L.~Freidel, R.~G. Leigh, and D.~Minic, ``{Quantum Spaces are Modular},''
  \href{https://dx.doi.org/10.1103/PhysRevD.94.104052}{{\em Phys. Rev.}
  {\bfseries D94} no.~10, (2016) 104052},
\href{https://arxiv.org/abs/1606.01829}{{\ttfamily arXiv:1606.01829 [hep-th]}}.
%%CITATION = ARXIV:1606.01829;%%.

\bibitem{Freidel:2017nhg}
L.~Freidel, R.~G. Leigh, and D.~Minic, ``{Noncommutativity of closed string
  zero modes},'' \href{https://dx.doi.org/10.1103/PhysRevD.96.066003}{{\em
  Phys. Rev.} {\bfseries D96} no.~6, (2017) 066003},
\href{https://arxiv.org/abs/1707.00312}{{\ttfamily arXiv:1707.00312 [hep-th]}}.
%%CITATION = ARXIV:1707.00312;%%.

\bibitem{Freidel:2017wst}
L.~Freidel, R.~G. Leigh, and D.~Minic, ``{Intrinsic non-commutativity of closed
  string theory},'' \href{https://dx.doi.org/10.1007/JHEP09(2017)060}{{\em J.
  High Energ. Phys.} {\bfseries 09} (2017) 060},
\href{https://arxiv.org/abs/1706.03305}{{\ttfamily arXiv:1706.03305 [hep-th]}}.
%%CITATION = ARXIV:1706.03305;%%.

\bibitem{Freidel:2017xsi}
L.~Freidel, R.~G. Leigh, and D.~Minic, ``{Modular Spacetime and Metastring
  Theory},''
\href{https://dx.doi.org/10.1088/1742-6596/804/1/012032}{{\em J. Phys. Conf.
  Ser.} {\bfseries 804} no.~1, (2017) 012032}.
%%CITATION = 00462,804,012032;%%.

\bibitem{Berglund:2020qcu}
P.~Berglund, T.~H\"ubsch, and D.~Minic, ``{String Theory, the Dark Sector and
  the Hierarchy Problem},''
  \href{https://dx.doi.org/10.31526/lhep.2021.186}{{\em LHEP} {\bfseries 2021}
  (2021) 186}, \href{https://arxiv.org/abs/2010.15610}{{\ttfamily
  arXiv:2010.15610 [hep-th]}}.

\bibitem{Freidel:2021wpl}
L.~Freidel, J.~Kowalski-Glikman, R.~G. Leigh, and D.~Minic, ``{Quantum Gravity
  Phenomenology in the Infrared},''
  \href{https://arxiv.org/abs/2104.00802}{{\ttfamily arXiv:2104.00802
  [gr-qc]}}.

\bibitem{Freidel:2022ryr}
L.~Freidel, J.~Kowalski-Glikman, R.~G. Leigh, and D.~Minic, ``{Vacuum energy
  density and gravitational entropy},''
  \href{https://dx.doi.org/10.1103/PhysRevD.107.126016}{{\em Phys. Rev. D}
  {\bfseries 107} no.~12, (2023) 126016},
  \href{https://arxiv.org/abs/2212.00901}{{\ttfamily arXiv:2212.00901
  [hep-th]}}.

\bibitem{Freidel:2023ytq}
L.~Freidel, J.~Kowalski-Glikman, R.~G. Leigh, and D.~Minic, ``{On the
  Inevitable Lightness of Vacuum},''
  \href{https://arxiv.org/abs/2303.17495}{{\ttfamily arXiv:2303.17495
  [hep-th]}}.

\bibitem{Minic:2023oty}
D.~Minic, ``The vacuum energy problem in quantum gravity and the masses of
  elementary particles,'' {\em Bulgarian J. Phys.} {\bfseries 50} (May, 2023)
  1--12, \href{https://arxiv.org/abs/2305.12593}{{\ttfamily arXiv:2305.12593
  [hep-th]}}.

\bibitem{Berglund:2023gur}
P.~Berglund, T.~H{\"u}bsch, and D.~Minic, ``String theory bounds on the
  cosmological constant, the {H}iggs mass, and the quark and lepton masses,''
  \href{https://dx.doi.org/10.3390/sym15091660}{{\em Symmetry} {\bfseries 15}
  no.~9, (8, 2023) 1660}, \href{https://arxiv.org/abs/2307.16712}{{\ttfamily
  arXiv:2307.16712 [hep-th]}}.

\bibitem{Hubsch:2026lot}
T.~H{\"u}bsch and D.~Minic, ``Quantum spacetime, quantum gravity and gravitized
  quantum theory,'' in {\em Nonlinearity, Nonlocality and Ultrametricity},
  Z.~Raki{\'c}, G.~Djordjevi{\'c}, and B.~Jovanovi{\'c}, eds.
\newblock Mathematical Institute of the Serbian Academy of Sciences and Arts,
  Belgrade, Serbia, 2026.
\newblock \href{https://arxiv.org/abs/2604.19418}{{\ttfamily arXiv:2604.19418
  [gr-qc]}}.

\bibitem{rF79a}
D.~H. Friedan, ``Nonlinear models in 2+$\epsilon$ dimensions,''
  \href{https://dx.doi.org/10.1103/PhysRevLett.45.1057}{{\em Phys. Rev. Lett.}
  {\bfseries 45} (1980) 1057}.

\bibitem{rF79b}
D.~H. Friedan, ``Nonlinear models in 2+$\epsilon$ dimensions,''
  \href{https://dx.doi.org/10.1016/0003-4916(85)90384-7}{{\em Ann. Phys.}
  {\bfseries 163} (1985) 318--419}.

\bibitem{Braaten:1985is}
E.~Braaten, T.~L. Curtright, and C.~K. Zachos, ``Torsion and geometrostasis in
  nonlinear sigma models,''
  \href{https://dx.doi.org/10.1016/0550-3213(85)90053-7}{{\em Nucl. Phys.}
  {\bfseries B260} (1985) 630}.

\bibitem{Berglund:2022qsb}
P.~Berglund, T.~H{\"u}bsch, and D.~Minic, ``On de~{S}itter spacetime and string
  theory,'' \href{https://dx.doi.org/10.1142/S0218271823300021}{{\em Int. J.
  Mod. Phys. D} {\bfseries 32} no.~9, (Dec., 2023) 2330002 (111 pages)},
  \href{https://arxiv.org/abs/2212.06086}{{\ttfamily arXiv:2212.06086
  [hep-th]}}.

\bibitem{Cheung:2025tbr}
C.~Cheung, G.~N. Remmen, F.~Sciotti, and M.~Tarquini, ``{Strings from Almost
  Nothing},'' \href{https://dx.doi.org/10.1103/cw4p-cqh7}{{\em Phys. Rev.
  Lett.} {\bfseries 136} no.~25, (2026) 251601},
  \href{https://arxiv.org/abs/2508.09246}{{\ttfamily arXiv:2508.09246
  [hep-th]}}.

\bibitem{Polchinski:2006gy}
J.~Polchinski, ``{The Cosmological Constant and the String Landscape},'' in
  {\em {23rd Solvay Conference in Physics: The Quantum Structure of Space and
  Time}}.
\newblock 3, 2006.
\newblock \href{https://arxiv.org/abs/hep-th/0603249}{{\ttfamily
  arXiv:hep-th/0603249}}.

\bibitem{Chang:2011jj}
L.~N. Chang, Z.~Lewis, D.~Minic, and T.~Takeuchi, ``{On the Minimal Length
  Uncertainty Relation and the Foundations of String Theory},''
  \href{https://dx.doi.org/10.1155/2011/493514}{{\em Adv. High Energy Phys.}
  {\bfseries 2011} (2011) 493514},
  \href{https://arxiv.org/abs/1106.0068}{{\ttfamily arXiv:1106.0068 [hep-th]}}.

\bibitem{Polchinski:1994mb}
J.~Polchinski, ``{What is string theory?},'' in {\em {NATO Advanced Study
  Institute: Les Houches Summer School, Session 62: Fluctuating Geometries in
  Statistical Mechanics and Field Theory}}.
\newblock 11, 1994.
\newblock \href{https://arxiv.org/abs/hep-th/9411028}{{\ttfamily
  arXiv:hep-th/9411028}}.

\bibitem{Ginsparg:1993is}
P.~H. Ginsparg and G.~W. Moore, ``{Lectures on 2-D gravity and 2-D string
  theory},'' in {\em {Theoretical Advanced Study Institute (TASI 92): From
  Black Holes and Strings to Particles}}, pp.~277--469.
\newblock 10, 1993.
\newblock \href{https://arxiv.org/abs/hep-th/9304011}{{\ttfamily
  arXiv:hep-th/9304011}}.

\bibitem{DiFrancesco:1993cyw}
P.~Di~Francesco, P.~H. Ginsparg, and J.~Zinn-Justin, ``{2-D Gravity and random
  matrices},'' \href{https://dx.doi.org/10.1016/0370-1573(94)00084-G}{{\em
  Phys. Rept.} {\bfseries 254} (1995) 1--133},
  \href{https://arxiv.org/abs/hep-th/9306153}{{\ttfamily
  arXiv:hep-th/9306153}}.

\bibitem{Taylor:2001vb}
W.~Taylor, ``{M(atrix) Theory: Matrix Quantum Mechanics as a Fundamental
  Theory},'' \href{https://dx.doi.org/10.1103/RevModPhys.73.419}{{\em Rev. Mod.
  Phys.} {\bfseries 73} (2001) 419--462},
  \href{https://arxiv.org/abs/hep-th/0101126}{{\ttfamily
  arXiv:hep-th/0101126}}.

\bibitem{Taylor:1999qk}
W.~Taylor, ``{The M(atrix) model of M theory},''
  \href{https://dx.doi.org/10.1007/978-94-011-4303-5_3}{{\em NATO Sci. Ser. C}
  {\bfseries 556} (2000) 91--178},
  \href{https://arxiv.org/abs/hep-th/0002016}{{\ttfamily
  arXiv:hep-th/0002016}}.

\bibitem{Maldacena:1997re}
J.~M. Maldacena, ``{The Large N limit of superconformal field theories and
  supergravity},'' \href{https://dx.doi.org/10.1023/A:1026654312961,
  10.4310/ATMP.1998.v2.n2.a1}{{\em Int. J. Theor. Phys.} {\bfseries 38} (1999)
  1113--1133}, \href{https://arxiv.org/abs/hep-th/9711200}{{\ttfamily
  arXiv:hep-th/9711200 [hep-th]}}.
[Adv. Theor. Math. Phys.2,231(1998)].
%%CITATION = HEP-TH/9711200;%%.

\bibitem{Gubser:1998bc}
S.~S. Gubser, I.~R. Klebanov, and A.~M. Polyakov, ``{Gauge theory correlators
  from noncritical string theory},''
  \href{https://dx.doi.org/10.1016/S0370-2693(98)00377-3}{{\em Phys. Lett.}
  {\bfseries B428} (1998) 105--114},
\href{https://arxiv.org/abs/hep-th/9802109}{{\ttfamily arXiv:hep-th/9802109
  [hep-th]}}.
%%CITATION = HEP-TH/9802109;%%.

\bibitem{Polchinski:1996fm}
J.~Polchinski, S.~Chaudhuri, and C.~V. Johnson, ``{Notes on D-branes},''
  \href{https://arxiv.org/abs/hep-th/9602052}{{\ttfamily
  arXiv:hep-th/9602052}}.

\bibitem{Casher:1985ra}
A.~Casher, F.~Englert, H.~Nicolai, and A.~Taormina, ``{Consistent Superstrings
  as Solutions of the D=26 Bosonic String Theory},''
  \href{https://dx.doi.org/10.1016/0370-2693(85)91072-X}{{\em Phys. Lett. B}
  {\bfseries 162} (1985) 121--126}.

\bibitem{Berglund:2021hbo}
P.~Berglund, T.~H\"ubsch, and D.~Minic, ``{Mirror Symmetry, Born Geometry and
  String Theory},'' in {\em {Nankai Symposium on Mathematical Dialogues}: {In
  celebration of S.-S.\:Chern's 110th anniversary}}, Y.-H. He, M.-L. Ge, C.-M.
  Bai, J.~Bao, and E.~Hirst, eds.
\newblock Springer Verlag, Singapore, Sept., 2022.
\newblock \href{https://arxiv.org/abs/2111.14205}{{\ttfamily arXiv:2111.14205
  [hep-th]}}.

\bibitem{DiVecchia:1987jb}
P.~Di~Vecchia, K.~Hornfeck, and M.~Yu, ``{Infinite Number of Conserved Charges
  and the $N$ String Vertex},''
  \href{https://dx.doi.org/10.1016/0370-2693(87)91573-5}{{\em Phys. Lett. B}
  {\bfseries 195} (1987) 557}.

\bibitem{rBeast2}
T.~H{\"u}bsch, {\em {C}alabi--{Y}au Manifolds: a Bestiary for Physicists}.
\newblock World Scientific Pub. Europe Ltd., London, UK, 2nd~ed., 2024.

\bibitem{Vafa:1989xc}
C.~Vafa, ``String vacua and orbifoldized {L-G} models,''
  \href{https://dx.doi.org/10.1142/S0217732389001350}{{\em Mod. Phys. Lett. A}
  {\bfseries 4} (1989) 1169}.

\bibitem{Intriligator:1990ua}
K.~A. Intriligator and C.~Vafa, ``Landau-{G}inzburg orbifolds,''
  \href{https://dx.doi.org/10.1016/0550-3213(90)90535-L}{{\em Nucl. Phys. B}
  {\bfseries 339} (1990) 95--120}.

\bibitem{Berglund:1991pp}
P.~Berglund and T.~Hubsch, ``{A Generalized construction of mirror
  manifolds},'' \href{https://dx.doi.org/10.1016/0550-3213(93)90250-S}{{\em
  Nucl. Phys. B} {\bfseries 393} (1993) 377--391},
  \href{https://arxiv.org/abs/hep-th/9201014}{{\ttfamily
  arXiv:hep-th/9201014}}.

\bibitem{Siegel:1993th}
W.~Siegel, ``{Superspace duality in low-energy superstrings},''
  \href{https://dx.doi.org/10.1103/PhysRevD.48.2826}{{\em Phys. Rev.}
  {\bfseries D48} (1993) 2826--2837},
\href{https://arxiv.org/abs/hep-th/9305073}{{\ttfamily arXiv:hep-th/9305073
  [hep-th]}}.
%%CITATION = HEP-TH/9305073;%%.

\bibitem{Siegel:1993xq}
W.~Siegel, ``{Two vierbein formalism for string inspired axionic gravity},''
  \href{https://dx.doi.org/10.1103/PhysRevD.47.5453}{{\em Phys. Rev.}
  {\bfseries D47} (1993) 5453--5459},
\href{https://arxiv.org/abs/hep-th/9302036}{{\ttfamily arXiv:hep-th/9302036
  [hep-th]}}.
%%CITATION = HEP-TH/9302036;%%.

\bibitem{Hull:1998vg}
C.~M. Hull, ``{Timelike T duality, de Sitter space, large N gauge theories and
  topological field theory},''
  \href{https://dx.doi.org/10.1088/1126-6708/1998/07/021}{{\em J. High Energ.
  Phys.} {\bfseries 07} (1998) 021},
\href{https://arxiv.org/abs/hep-th/9806146}{{\ttfamily arXiv:hep-th/9806146
  [hep-th]}}.
%%CITATION = HEP-TH/9806146;%%.

\bibitem{Hull:2004in}
C.~M. Hull, ``{A Geometry for non-geometric string backgrounds},''
  \href{https://dx.doi.org/10.1088/1126-6708/2005/10/065}{{\em JHEP} {\bfseries
  10} (2005) 065}, \href{https://arxiv.org/abs/hep-th/0406102}{{\ttfamily
  arXiv:hep-th/0406102}}.

\bibitem{Hull:2009mi}
C.~Hull and B.~Zwiebach, ``{Double Field Theory},''
  \href{https://dx.doi.org/10.1088/1126-6708/2009/09/099}{{\em J. High Energ.
  Phys.} {\bfseries 09} (2009) 099},
\href{https://arxiv.org/abs/0904.4664}{{\ttfamily arXiv:0904.4664 [hep-th]}}.
%%CITATION = ARXIV:0904.4664;%%.

\bibitem{Gunaydin:2013nqa}
M.~G{\"u}naydin and D.~Minic, ``{Nonassociativity, Malcev Algebras and String
  Theory},'' \href{https://dx.doi.org/10.1002/prop.201300010}{{\em Fortsch.
  Phys.} {\bfseries 61} (2013) 873--892},
  \href{https://arxiv.org/abs/1304.0410}{{\ttfamily arXiv:1304.0410 [hep-th]}}.

\bibitem{Gunaydin:1973rs}
M.~G{\"u}naydin and F.~G{\"u}rsey, ``{Quark structure and octonions},''
  \href{https://dx.doi.org/10.1063/1.1666240}{{\em J. Math. Phys.} {\bfseries
  14} (1973) 1651--1667}.

\bibitem{Gunaydin:1974fb}
M.~G{\"u}naydin and F.~G{\"u}rsey, ``{Quark Statistics and Octonions},''
  \href{https://dx.doi.org/10.1103/PhysRevD.9.3387}{{\em Phys. Rev. D}
  {\bfseries 9} (1974) 3387}.

\bibitem{Gursey:1996mj}
F.~G{\"u}rsey and H.~C. Tze, {\em {On the role of division, Jordan and related
  algebras in particle physics}}.
\newblock World Scientific, 1996.

\bibitem{Aharonov:2005uc}
Y.~Aharonov and D.~Rohrlich, {\em Quantum Paradoxes: Quantum Theory for the
  Perplexed}.
\newblock Wiley-VCH, 2005.

\bibitem{Amelino-Camelia:2011hjg}
G.~Amelino-Camelia, L.~Freidel, J.~Kowalski-Glikman, and L.~Smolin, ``{Relative
  locality: A deepening of the relativity principle},''
  \href{https://dx.doi.org/10.1142/S0218271811020743}{{\em Gen. Rel. Grav.}
  {\bfseries 43} (2011) 2547--2553},
  \href{https://arxiv.org/abs/1106.0313}{{\ttfamily arXiv:1106.0313 [hep-th]}}.

\bibitem{Amelino-Camelia:2011lvm}
G.~Amelino-Camelia, L.~Freidel, J.~Kowalski-Glikman, and L.~Smolin, ``Principle
  of relative locality,''
  \href{https://dx.doi.org/10.1103/PhysRevD.84.084010}{{\em Phys. Rev. D}
  {\bfseries 84} (2011) 084010},
  \href{https://arxiv.org/abs/1101.0931}{{\ttfamily arXiv:1101.0931 [hep-th]}}.

\bibitem{Freidel:2018apz}
L.~Freidel, J.~Kowalski-Glikman, R.~G. Leigh, and D.~Minic, ``{Theory of
  metaparticles},'' \href{https://dx.doi.org/10.1103/PhysRevD.99.066011}{{\em
  Phys. Rev. D} {\bfseries 99} no.~6, (2019) 066011},
  \href{https://arxiv.org/abs/1812.10821}{{\ttfamily arXiv:1812.10821
  [hep-th]}}.

\bibitem{Freidel:2019jor}
L.~Freidel, R.~G. Leigh, and D.~Minic, ``{Manifest non-locality in quantum
  mechanics, quantum field theory and quantum gravity},''
\href{https://dx.doi.org/10.1142/9789811213144_0004,
  10.1142/S0217751X19410045}{{\em Int. J. Mod. Phys.} {\bfseries A34} no.~28,
  (2019) 1941004}.
%%CITATION = IMPAE,A34,1941004;%%.

\bibitem{Grosse:2004yu}
H.~Grosse and R.~Wulkenhaar, ``{Renormalization of phi**4 theory on
  noncommutative R**4 in the matrix base},''
  \href{https://dx.doi.org/10.1007/s00220-004-1285-2}{{\em Commun. Math. Phys.}
  {\bfseries 256} (2005) 305--374},
\href{https://arxiv.org/abs/hep-th/0401128}{{\ttfamily arXiv:hep-th/0401128
  [hep-th]}}.
%%CITATION = HEP-TH/0401128;%%.

\bibitem{Moore:1993zc}
G.~W. Moore, ``{Finite in all directions},''
  \href{https://arxiv.org/abs/hep-th/9305139}{{\ttfamily
  arXiv:hep-th/9305139}}.

\bibitem{Gopakumar:1994iq}
R.~Gopakumar and D.~J. Gross, ``{Mastering the master field},''
  \href{https://dx.doi.org/10.1016/0550-3213(95)00340-X}{{\em Nucl. Phys. B}
  {\bfseries 451} (1995) 379--415},
  \href{https://arxiv.org/abs/hep-th/9411021}{{\ttfamily
  arXiv:hep-th/9411021}}.

\bibitem{Penrose:2015lla}
R.~Penrose, ``{Palatial twistor theory and the twistor googly problem},''
  \href{https://dx.doi.org/10.1098/rsta.2014.0237}{{\em Phil. Trans. Roy. Soc.
  Lond. A} {\bfseries 373} (2015) 20140237}.

\bibitem{Atick:1988si}
J.~J. Atick and E.~Witten, ``{The Hagedorn Transition and the Number of Degrees
  of Freedom of String Theory},''
  \href{https://dx.doi.org/10.1016/0550-3213(88)90151-4}{{\em Nucl. Phys. B}
  {\bfseries 310} (1988) 291--334}.

\bibitem{Sen:2024nfd}
A.~Sen and B.~Zwiebach, ``{String Field Theory: A Review},''
  \href{https://arxiv.org/abs/2405.19421}{{\ttfamily arXiv:2405.19421
  [hep-th]}}.

\bibitem{Witten:1985cc}
E.~Witten, ``{Noncommutative Geometry and String Field Theory},''
  \href{https://dx.doi.org/10.1016/0550-3213(86)90155-0}{{\em Nucl. Phys. B}
  {\bfseries 268} (1986) 253--294}.

\bibitem{Strominger:1987ad}
A.~Strominger, ``Closed string field theory,''
  \href{https://dx.doi.org/10.1016/0550-3213(87)90574-8}{{\em Nucl. Phys. B}
  {\bfseries 294} (1987) 93--112}.

\bibitem{Ashtekar:1997ud}
A.~Ashtekar and T.~A. Schilling, ``{Geometrical formulation of quantum
  mechanics},'' \href{https://arxiv.org/abs/gr-qc/9706069}{{\ttfamily
  arXiv:gr-qc/9706069}}.

\bibitem{Wootters:1981ki}
W.~K. Wootters, ``{Statistical Distance and Hilbert Space},''
  \href{https://dx.doi.org/10.1103/PhysRevD.23.357}{{\em Phys. Rev. D}
  {\bfseries 23} (1981) 357--362}.

\bibitem{Aharonov:1987Pha}
Y.~Aharonov and J.~Anandan, ``Phase change during a cyclic quantum evolution,''
  \href{https://dx.doi.org/10.1103/PhysRevLett.58.1593}{{\em Phys. Rev. Lett.}
  {\bfseries 58} (Apr, 1987) 1593--1596}.

\bibitem{Dakic:2014ntj}
B.~Dakic, T.~Paterek, and C.~Brukner, ``{Density cubes and higher-order
  interference theories},''
  \href{https://dx.doi.org/10.1088/1367-2630/16/2/023028}{{\em New J. Phys.}
  {\bfseries 16} (2014) 023028},
  \href{https://arxiv.org/abs/1308.2822}{{\ttfamily arXiv:1308.2822
  [quant-ph]}}.

\bibitem{Namdar:2021czo}
P.~Namdar, P.~K. Jenke, I.~A. Calafell, A.~Trenti, M.~Radonji\'c, B.~Daki\'c,
  P.~Walther, and L.~A. Rozema, ``Experimental higher-order interference in a
  nonlinear triple slit,''
  \href{https://dx.doi.org/10.1103/PhysRevA.107.032211}{{\em Phys. Rev. A}
  {\bfseries 107} no.~3, (2023) 032211},
  \href{https://arxiv.org/abs/2112.06965}{{\ttfamily arXiv:2112.06965
  [quant-ph]}}.

\bibitem{Brezin:1992yc}
E.~Brezin and J.~Zinn-Justin, ``{Renormalization group approach to matrix
  models},'' \href{https://dx.doi.org/10.1016/0370-2693(92)91953-7}{{\em Phys.
  Lett. B} {\bfseries 288} (1992) 54--58},
  \href{https://arxiv.org/abs/hep-th/9206035}{{\ttfamily
  arXiv:hep-th/9206035}}.

\bibitem{Hagston:1980xe}
W.~E. Hagston and M.~Roberts, ``Formal manipulations relating the s matrix and
  density matrix formalisms,''
  \href{https://dx.doi.org/10.1088/0305-4470/13/8/019}{{\em J. Phys. A}
  {\bfseries 13} (1980) 2695--2702}.

\bibitem{Pagani:2022pjq}
C.~Pagani and M.~Reuter, ``{Time evolution of density matrices as a theory of
  random surfaces},'' \href{https://dx.doi.org/10.1016/j.aop.2023.169361}{{\em
  Annals Phys.} {\bfseries 455} (2023) 169361},
  \href{https://arxiv.org/abs/2212.13748}{{\ttfamily arXiv:2212.13748
  [hep-th]}}.

\bibitem{Edmonds:2024qsj}
D.~Edmonds, J.~Erlich, D.~Minic, and T.~Takeuchi, ``Universal acceleration and
  fuzzy dark matter,'' \href{https://arxiv.org/abs/2405.08744}{{\ttfamily
  arXiv:2405.08744 [astro-ph.CO]}}.

\bibitem{Berglund:2019ctg}
P.~Berglund, T.~H{\"u}bsch, and D.~Minic, ``{Dark Energy and String Theory},''
  \href{https://dx.doi.org/10.1016/j.physletb.2019.134950}{{\em Phys. Lett. B}
  {\bfseries 798} (2019) 134950},
  \href{https://arxiv.org/abs/1905.08269}{{\ttfamily arXiv:1905.08269
  [hep-th]}}.

\bibitem{Hur:2025lqc}
S.~Hur, V.~Jejjala, M.~J. Kavic, D.~Minic, and T.~Takeuchi, ``Dynamical dark
  energy, dual spacetime, and {DESI},''
  \href{https://dx.doi.org/10.1088/1475-7516/2025/11/054}{{\em JCAP} {\bfseries
  11} (2025) 054}, \href{https://arxiv.org/abs/2503.20854}{{\ttfamily
  arXiv:2503.20854 [hep-th]}}.

\end{thebibliography}
\providecommand{\href}[2]{#2}\begingroup\raggedright\endgroup

\endgroup

\end{document}